\RequirePackage{fix-cm}
\documentclass[smallextended]{svjour3}       
\smartqed  
\usepackage[utf8]{inputenc}
\usepackage{graphicx}
\usepackage{graphics}
\usepackage[round,sort]{natbib}
\usepackage{hhline}
\usepackage{comment}
\usepackage[dvipsnames]{xcolor}
\usepackage[colorlinks=false,linkcolor=blue,citecolor=blue,bookmarks=true,pagebackref=false,hidelinks=true]{hyperref} 
\usepackage{hyphenat}
\usepackage{subcaption}
\usepackage{float}
\usepackage{caption}
\usepackage{fancybox}
\usepackage{array,booktabs}
\usepackage{marvosym} 
\usepackage{textcomp} 
\usepackage{tablefootnote} 
\usepackage{balance}
\usepackage{multirow}
\usepackage{soul} 
\usepackage[flushleft]{threeparttable} 
\usepackage{enumitem} 
\usepackage{tcolorbox} 
\usepackage{listings, lstautogobble} 
\usepackage{url}
\usepackage{adjustbox}
\usepackage{pdfpages}
\usepackage{commath}
\usepackage{amsmath}

\definecolor{mycolor}{RGB}{10, 186, 181}

\newcommand{\phead}[1]{\vspace{1mm} \noindent {\bf #1}}
\newcommand{\uhead}[1]{\vspace{1mm} \noindent {\underline {#1}}}

\definecolor{dkgreen}{rgb}{0,0.6,0}
\definecolor{gray}{rgb}{0.5,0.5,0.5}
\definecolor{mauve}{rgb}{0.58,0,0.82}

\lstdefinestyle{customeStackTracesStyle}{
  frame=none,
  language=Java,
  showstringspaces=false,
  columns=flexible,
  basicstyle={\scriptsize\ttfamily},
  numbers=none,
  numberstyle=\tiny\color{gray},
  keywordstyle=\color{blue},
  commentstyle=\color{dkgreen},
  stringstyle=\color{mauve},
  breaklines=true,
  breakatwhitespace=true,
  tabsize=1,
  keywordsprefix={at}
}

\usepackage{makecell}
\usepackage{courier}
\lstdefinestyle{customeLogSnippets}{
  frame=none,
  language=Java,
  showstringspaces=false,
  columns=flexible,
  basicstyle={\scriptsize\ttfamily},
  numbers=none,
  numberstyle=\tiny\color{gray},
  keywordstyle=\color{blue},
  commentstyle=\color{dkgreen},
  stringstyle=\color{mauve},
  breaklines=true,
  breakatwhitespace=true,
  tabsize=1,
  keywordsprefix={INFO}
}

\lstdefinestyle{customeLogSnippetsWarn}{
  frame=none,
  language=Java,
  showstringspaces=false,
  columns=flexible,
  basicstyle={\scriptsize\ttfamily},
  numbers=none,
  numberstyle=\tiny\color{gray},
  keywordstyle=\color{blue},
  commentstyle=\color{dkgreen},
  stringstyle=\color{mauve},
  breaklines=true,
  breakatwhitespace=true,
  tabsize=1,
  keywordsprefix={WARN}
}

\lstdefinestyle{customeJavaStyle}{
  language=Java,
  tabsize=4,
  showspaces=false,
  showstringspaces=false,
  numberstyle=\footnotesize,
}

\lstdefinestyle{customeJavaStyleWithLine}{
  language=Java,
  numbers=left,
  stepnumber=1,
  numbersep=10pt,
  tabsize=4,
  showspaces=false,
  showstringspaces=false,
  numberstyle=\footnotesize,
}

\lstdefinestyle{base}{
  breaklines=true,
  basicstyle=\ttfamily\color{black},
  moredelim=**[is][\color{red}]{@}{@},
}

\makeatletter

\begin{document}


\title{SBEST: Spectrum-Based Fault Localization Without Fault-Triggering Tests}
\titlerunning{SBEST: Spectrum-Based Fault Localization Without Fault-Triggering Tests}

\author{
Md Nakhla Rafi\textsuperscript{*}, Lorena Barreto Simedo Pacheco\textsuperscript{*} \and 
An Ran Chen \and Jinqiu Yang \and Tse-Hsun (Peter) Chen
}

\institute{
Md Nakhla Rafi\textsuperscript{*}, Lorena Barreto Simedo Pacheco\textsuperscript{*}, Jinqiu Yang, Tse-Hsun (Peter) Chen \at
Concordia University, Montreal, Canada\\
\email{lorena.bspacheco@mail.concordia.ca}, \email{\{jinqiuy, peterc\}@encs.concordia.ca}
\and
An Ran Chen \at University of Alberta, Edmonton, Canada \\
\email{anran6@ualberta.ca}
}

\footnotetext[1]{* These authors contributed equally to this work.}

\date{Received: date / Accepted: date}

\maketitle

\begin{abstract}
Fault localization is a critical step in software maintenance. Yet, many existing techniques, such as Spectrum-Based Fault Localization (SBFL), rely heavily on the availability of fault-triggering tests to be effective. In practice, especially for crash-related bugs, such tests are frequently unavailable. Meanwhile, bug reports containing stack traces often serve as the only available evidence of runtime failures and provide valuable context for debugging. This study investigates the feasibility of using stack traces from crash reports as proxies for fault-triggering tests in SBFL. Our empirical analysis of 60 crash-report bugs in Defects4J reveals that only 3.33\% of these bugs have fault-triggering tests available at the time of the bug report creation. However, 98.3\% of bug fixes directly address the exception observed in the stack trace, and 78.3\% of buggy methods are reachable within an average of 0.34 method calls from the stack trace. These findings underscore the diagnostic value of stack traces in the absence of failing tests. Motivated by these findings, we propose SBEST, a novel approach that integrates stack trace information with test coverage data to perform fault localization when fault-triggering tests are missing. SBEST shows an improvement, with a 32.22\% increase in Mean Average Precision (MAP) and a 17.43\% increase in Mean Reciprocal Rank (MRR) compared to baseline approaches under the scenario where fault-triggering tests are absent.


\end{abstract}

\section{Introduction}\label{sec:introduction}
Locating and fixing software bugs is an essential but often expensive and time-consuming task. Bug-fixing activities delay project progress and increase overall maintenance costs. As modern software systems grow in size and complexity, efficiently and accurately localizing faults has become increasingly critical.
To address this challenge, previous research has proposed numerous different automated fault localization techniques \citep{abreu-etal-2007-accuracy, jones-harrold-2005-tarantula, naish2011model, abreu2009spectrum, wong2013dstar, papadakis2015metallaxis,hong2017museum, papadakis2012using, 9759520, 9201269}. Among these, Spectrum-Based Fault Localization (SBFL) stands out due to its conceptual simplicity, low computational cost, and exemplary performance in many scenarios \citep{de2016spectrum}. SBFL techniques leverage information derived from test executions, specifically the spectrum of code elements executed during passing and failing tests, to calculate the probability of every code element being faulty (i.e., suspiciousness score).

Although SBFL is no longer considered state-of-the-art in terms of absolute accuracy, it remains highly relevant and widely adopted today. For example, prior learning-based fault localization techniques~\citep{li2019deepfl,sohn2017fluccs} use SBFL scores as core input features for model training. More recently, LLM-based approaches like AutoCodeRover \citep{zhang2024autocoderover} and LLAMAO \citep{yang2024large} also rely on SBFL outputs to guide code analysis and method ranking. Beyond fault localization, SBFL plays a central role in various downstream software engineering tasks, including test case prioritization~\citep{li2007search,srivastava2008test}, and automated program repair (APR)~\citep{motwani2023better, kuma2020improving, lou2020can}. 


The effectiveness of SBFL highly depends on the availability of failing test cases that uncover the bug, known as fault-triggering tests~\citep{chen-etal-2022-useful}. A fault-triggering test is a test case that fails specifically because it executes the buggy code responsible for the fault, and passes once the bug is fixed. However, studies indicate that in real-world scenarios, especially in continuous deployment and integration environments, such tests may not always be present or identified when bugs are reported \citep{chen2023back, kabadi2023future}. The absence of fault-triggering tests can significantly impact the applicability of SBFL, as it relies on them to distinguish between faulty and non-faulty code.

\begin{figure*}
    \centering
\includegraphics[scale=0.3]{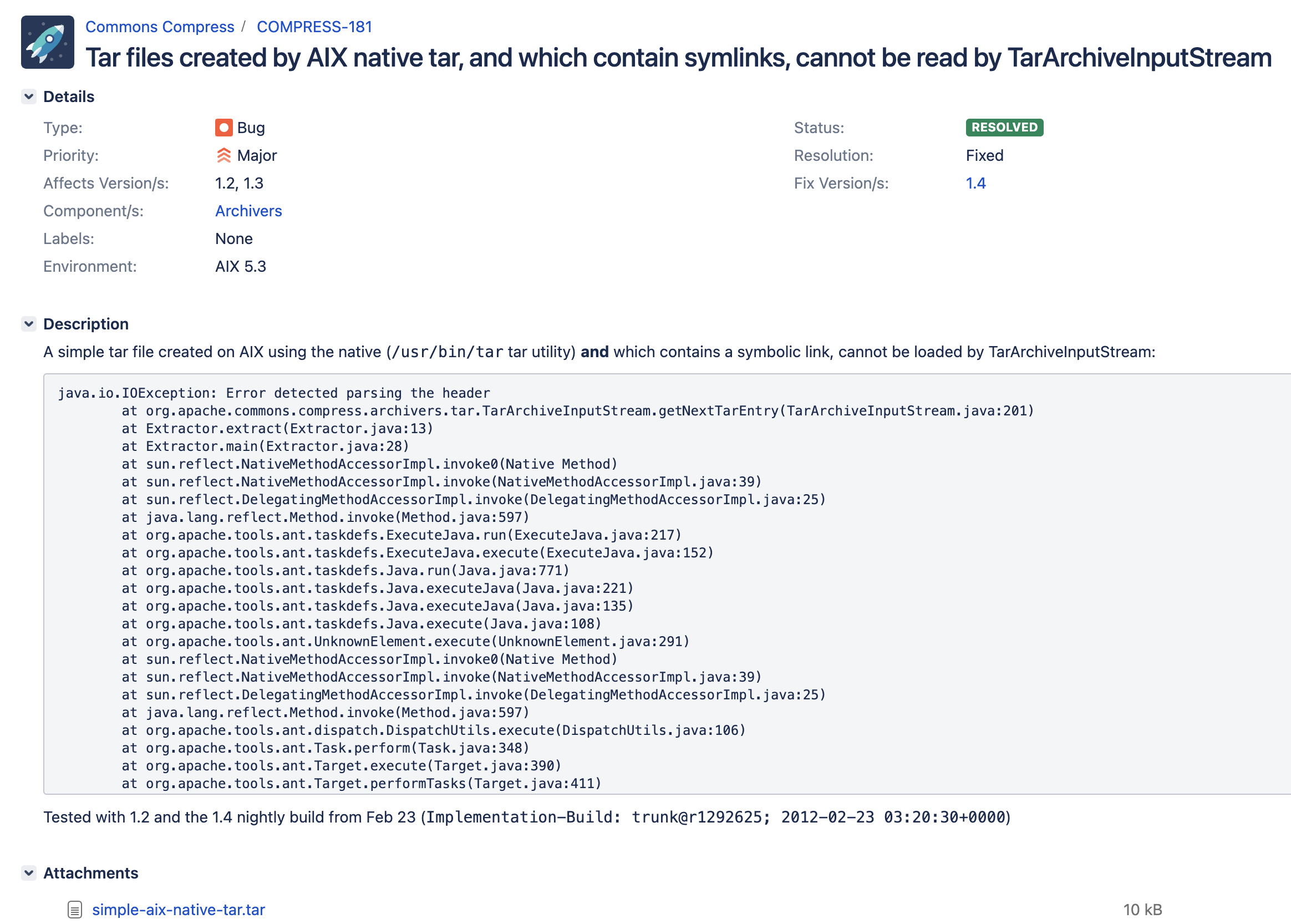}
    \caption{Example bug report (COMPRESS-181). The bug report addresses a problem in reading TAR files that contain symlinks, which cause an IOException to be thrown. The report contains the stack trace of the error (gray box), and an example of TAR file (simple-aix-native-tar.tar) that triggers the bug is attached. When this bug report was created, there were no failing tests.}
    \label{fig:bug_report_motivation}
\end{figure*}

In such situations, developers often rely on \textit{Crash Reports}, a specific type of bug report that documents unexpected program terminations along with diagnostic details such as exception messages and stack traces. A crash report typically includes a title, description, and the stack trace that captures the program’s execution path leading to the failure \citep{chen-etal-2021-demystifying}. For example, Figure \ref{fig:bug_report_motivation} shows a crash report from the Defects4J 2.0 dataset (COMPRESS-181), which describes an IOException triggered when reading TAR files with symbolic links. The report includes the relevant stack trace but no fault-triggering tests at the time, making SBFL unable to prioritize the faulty locations. This example highlights how stack trace sequences of method calls preceding an exception can serve as an alternative source of execution information when traditional test-based approaches are not applicable \citep{schroter2010stack, wu2014crashlocator}.

\par
In this paper, we address the challenges of fault localization in scenarios where traditional fault-triggering test data is scarce. We conduct our study on the Defects4J 2.0 dataset \citep{just-etal-2014-defects4j}, a widely recognized benchmark for research in fault localization and program repair. 
Prior studies \citep{kabadi2023future, chen2023back} report that some Defects4J tests were added post hoc, potentially biasing localization. To avoid this, we analyze the commit preceding the bug report to better reflect the state of the code repository that developers saw at the time. 
Our initial analysis of the bugs reveals that only 3.33\% of crash reports, which are bug reports that contain stack traces, possess fault-triggering tests. Even for bug reports without stack traces, only around 10\% of the bugs have fault-triggering tests. This low incidence of failing tests challenges the efficacy of Spectrum-Based Fault Localization (SBFL) methods, which rely heavily on test outcomes to identify bug locations. To overcome this limitation, we propose leveraging stack traces, which are rich in contextual and execution information about the bug’s root causes, as a substitute for the failing tests. We introduce a novel approach, Spectrum-Based Localization Enhanced by Stack Traces (SBEST), which integrates stack trace analysis within the SBFL framework. Our method improves fault localization accuracy in scenarios where fault-triggering tests are unavailable by combining coverage information with the execution context provided by exception stack traces. This integration enables a more detailed understanding of system behaviour at the moment of failure, thereby addressing the gaps left by the unavailability of failing tests.

In particular, we study and answer three research questions (RQs):

\begin{itemize}
    \item {\textbf{RQ1: Are the test failures related to the bug in crash reports?} We examine whether the test failures observed at the bug report commit are truly fault-triggering, that is, caused by the bug described in the crash report. Our analysis shows that only 3.33\% of crash-report bugs have such fault-triggering tests, limiting the applicability of SBFL. This finding underscores the need to augment SBFL with alternative sources of execution information, such as stack traces, especially when traditional failing tests are unavailable.}
    
    \item {\textbf{RQ2: What is the relationship between stack traces and buggy location?} Given that the stack traces are the only execution information available in most of the studied bug reports, we aim to study how they relate to the buggy location and how they can be used to locate these bugs. To do so, we inspect the type of modification performed in the bug fix commit and the distance between the stack trace and the buggy methods. We find that in 98.3\% of the studied bugs, the bugfix intention is directly correlated with the exception in the stack trace (e.g., adding code to handle the exception). In addition, 78.3\% of the buggy methods are directly reachable from the stack traces, with an average distance of 0.34 method calls. This indicates that leveraging the stack traces to reconstruct the execution path at the time of the exception can be an effective method for pinpointing bugs. } 
    \item {\textbf{RQ3: Can we utilize the stack traces to help detect the buggy locations?} We propose SBEST, a technique that leverages stack traces as a proxy for fault-triggering tests within the SBFL framework. To evaluate SBEST, we compare it against four baselines: Ochiai, Stack Trace ranking, SB\_score (stack trace + Ochiai), and Call Frequency-Based Fault Localization. Among the baselines, Stack Trace alone ranks 34 out of 60 bugs in the Top-5, while Ochiai identifies only 2. SBEST localizes 33 bugs in the Top-5, with MAP and MRR values 32.2\% and 17.4\% higher than the Stack Trace baseline, respectively. }

\end{itemize}

In summary, this paper presents a comprehensive study evaluating the effectiveness of combining stack traces with test coverage data. Our findings indicate that incorporating stack traces into the SBFL model can enhance bug localization in scenarios where fault-triggering tests are unavailable, addressing an important challenge in automated debugging. We make the data publicly available online for replication and encourage future research in this area \citep{sbest}.

\noindent\textbf{Paper Organization.}
Section \ref{sec:related} surveys the related work. Section \ref{sec:case} explains the experimental setup. Section \ref{sec:rqs} presents the results. Section \ref{sec:discussion} discusses the implications of our findings. Section \ref{sec:threat} presents the threats to validity. Section \ref{sec:conclusion} concludes the paper and discusses future work.

\section{Related Work}\label{sec:related}

\subsection{Fault Localization}
Automated fault localization methods are a recurrent theme in the literature, being Spectrum-Based Fault Localization (SBFL) and Mutation-Based Fault Localization (MBFL) being two of the most famous representatives. 
SBFL is a very consolidated class of Fault Localization techniques. These approaches utilize test information to localize bugs, and they require at least one failing test to be effective. Ochiai \citep{abreu-etal-2007-accuracy} is one of the most famous SBFL techniques. Other representatives of this class are Tarantula \citep{jones-harrold-2005-tarantula}, Op2 \citep{naish2011model}, BARINEL \citep{abreu2009spectrum}, and DStar \citep{wong2013dstar}: similar approaches based on different formulas to compute the suspiciousness score. Given that SBFL techniques are lightweight and fast, they are also applied to other types of methods, such as interactive fault localization \citep{gong2012interactive} and program repair \citep{ye2022neural, le2012systematic}. 

Information Retrieval-based Fault Localization (IRFL) approaches aim to localize buggy code by treating bug reports as textual queries and comparing them against the source code using similarity metrics. Early techniques, such as BugLocator~\citep{zhou2012should}, enhanced traditional IR techniques by combining TF-IDF-based scoring with historical bug-fix data to improve the ranking of relevant files. Subsequent work, such as BLUiR~\citep{saha2013improving}, further advanced this by leveraging structured information from source code, including method and class names, to improve retrieval accuracy over treating code as flat text. In addition, BRTracer~\citep{wong2014boosting} improved IR-based fault localization by segmenting source files and boosting files referenced in stack traces. Additionally, hybrid approaches, such as the one proposed by~\cite{lam2017bug}, combine deep learning (e.g., CNNs) with IR signals to better capture the semantic relationships between bug reports and source code, outperforming purely IR-based baselines. Supervised IR-based models such as STMLocator+~\citep{wang2020enhancing} further enhanced accuracy by integrating textual and semantic similarity, metadata, and stack trace information.

MBFL techniques~\citep{papadakis2015metallaxis,moon2014ask,hong2017museum,papadakis2012using,9759520} are based on mutation analysis. For each part of the code covered by failed test cases, a set of mutants is created, and the test suite is executed. Based on the test results, the suspiciousness scores are calculated, and the most suspicious locations are ranked. Despite presenting promising results, this kind of technique is very costly and time-consuming since it requires multiple executions of the test suite. Strategies for reducing this overhead and improving the general accuracy have been studied, such as ways of prioritizing the mutants \citep{liu2018optimal}, prioritizing the tests executed for each mutant \citep{de2018ftmes}, and a mix of spectrum and mutant-based approaches \citep{9201269}. One common point between SBFL and MBFL approaches is that both types of techniques require failing tests to work. However, this might not always be the case in real life.

Call Frequency-Based Fault Localization (CFBFL)~\citep{callfreq} is another dynamic approach that does not rely on coverage spectra but instead ranks methods based on how frequently they are executed across tests. Unlike simple count-based methods, CFBFL considers the number of times a method appears in unique call stack contexts during failing executions, filtering out noise from loops and repetitive calls. A related approach by ~\cite{levcontextfl} improves SBFL by analyzing function call chains, computing suspiciousness over call paths, and then attributing it to individual functions. Both techniques enhance fault localization by leveraging richer execution context beyond binary coverage.

Most of the mentioned studies use Defects4J to evaluate the results. Defects4J is one of the most widely used benchmarks for fault localization techniques that rely on test coverage to identify suspicious code elements. Nonetheless, recent studies \citep{kabadi2023future, chen2023back} show that often in Defects4J, tests created after the bug report are added to the buggy patch and thus contain developer knowledge about the bug investigation process. These tests were artificially appended to the buggy commit in Defects4J to simulate a scenario with a bug and test failures. This fact suggests that many fault localization techniques evaluated in Defects4J may yield significantly different results when applied to real-world scenarios.

\subsection{Bug Reports and Stack Traces}
Bug report quality has long been a central topic in software engineering research. Understanding which elements of a bug report are most helpful in identifying and resolving faults can help guide the development of more effective fault localization techniques. One of the most consistently valuable components of bug reports is the stack trace, particularly in the context of crashes and runtime errors.

\citet{bettenburg2008makes} conducted a developer survey to determine what makes a bug a helpful report. Their results highlight the importance of detailed reproduction steps, stack traces, and associated test cases. These elements help developers quickly understand and reproduce the issue, making stack traces especially valuable for debugging. \citet{schroter2010stack} further investigated the connection between stack traces and the actual faults by analyzing whether methods in the reported stack traces were modified during the corresponding bug fix. Their findings revealed that in over 47\% of the cases, at least one buggy method appeared in the stack trace, suggesting that stack traces often point directly to fault-relevant areas of the code.

Building on this, \cite{chen-etal-2021-demystifying} performed a large-scale empirical study of 1,561 bug reports containing logs, including stack traces and log messages. They found that in 73\% of the cases, there was an overlap between the classes mentioned in the logs and the classes involved in the bug fix. This reinforces the view that logs and especially stack traces provide strong signals for localizing faults.
Several techniques have been proposed that leverage stack traces more directly for automated fault localization. For example,~\cite{jiang2012stack} use stack traces in conjunction with program slicing to localize null pointer exceptions better. \cite{gong2014locating} generates stack traces to aid in crash fault localization and demonstrates their approach’s effectiveness by identifying 64\% of crashing faults in Firefox 3.6.


CrashLocator~\citep{wu2014crashlocator} introduced a scalable method for leveraging stack traces across multiple crash reports. It clusters similar crash reports into buckets corresponding to the same fault and cross-analyzes the slightly varying stack traces within each bucket. The approach expands stack traces using static call graph traversal and ranks methods based on their frequency within the cluster, rarity across other clusters, and proximity to the reported crash point. This design allows CrashLocator to combine statistical signals with structural code context, yielding highly accurate fault predictions even in large-scale systems. Together, these studies and tools highlight the value of stack traces not only as debugging aids for developers but also as key artifacts that can be mined and augmented for automated fault localization.

\noindent\textbf{Comparison with Closely Related Studies.}
While prior studies have demonstrated the usefulness of stack traces for fault localization, their objectives and assumptions differ from ours. 
CrashLocator~\citep{wu2014crashlocator} clusters multiple crash reports and expands stack traces using static call graphs to rank methods by their frequency and proximity within crash clusters. 
In contrast, SBEST focuses on individual crash reports and integrates stack-trace-derived execution paths directly into the Spectrum-Based Fault Localization (SBFL) framework, treating stack-trace methods as proxies for fault-triggering tests. 
Call Frequency-Based Fault Localization (CFBFL)~\citep{callfreq} extends Spectrum-Based Fault Localization by weighting coverage information with the frequency of method occurrences in stack traces collected during test executions.
Information Retrieval (IR)-based techniques such as BugLocator~\citep{zhou2012should} and STMLocator+~\citep{wang2020enhancing} rely mainly on textual similarity between bug reports and source code, occasionally incorporating stack traces as supplementary input.
In contrast, SBEST leverages stack traces extracted from crash reports rather than from executed tests and integrates them with coverage data to perform fault localization even when fault-triggering tests are missing.

\section{Data Collection and Study Design}\label{sec:case}

In this section, we start by describing the data collection process and presenting our dataset. After this, we detail the evaluation metrics. 

\subsection{Data Collection}

\subsubsection{Collecting Bug Reports with Stack Traces}\label{sec:test_exection}
In this paper, we study crash reports (i.e., bug reports that contain stack traces) in 15 projects from Defects4J version 2.0.0. Defects4J \citep{just-etal-2014-defects4j} is a benchmark and framework of real bugs used in many previous Software Engineering studies related to bug repair and fault localization \citep{lutellier-etal-2020-coconut, chen-etal-2021-pathidea, li2021fault, ye2022neural}. 
The bugs in Defects4j comprise a wide variety of systems with different characteristics, with sizes ranging from 4K to 90K lines of code (LOC) and the number of tests varying from 54 to 7,911. 
Although Defects4J provides some basic information about the bugs (e.g., bug report ID and the commit hash of the fixes), it does not contain the stack traces provided in the bug reports. Hence, we start by extracting the bug report URL for all the bugs in the Defects4J repository. To collect the bug reports' textual information (title, description, and comments) and their creation date, we implement a crawler to retrieve information from the project management tool of each system (Jira, GitHub Issues, or Source Forge). 

\begin{figure}
\includegraphics[scale=0.25]{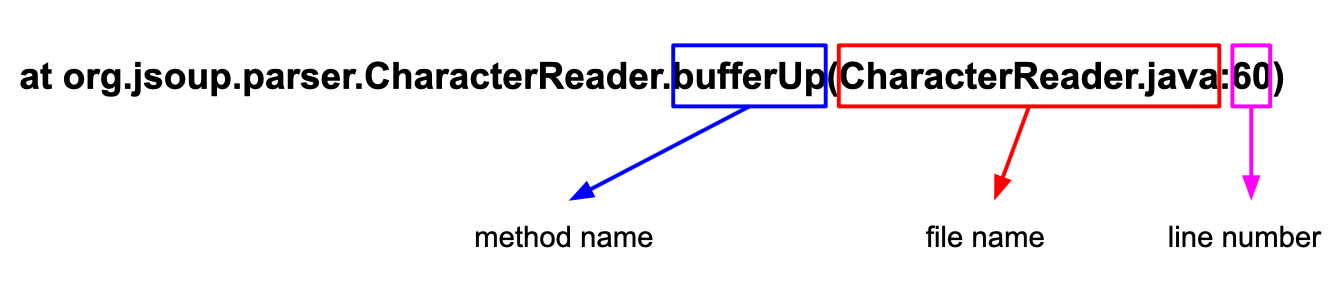}
\caption{\label{fig:stackTracesStructure}  An example of the structure of stack trace entries in Java. }
\end{figure}
\par
In total, we collected 803 bugs for which the corresponding bug reports were available.  For each bug, we then combine the bug report information with the information available in Defects4J, including the bugfix commit hash and the list of fault-triggering tests. \textit{\textbf{Fault-triggering tests are defined as tests that fail in the buggy commit and pass in the bugfix commit}}. They are tests that cover the buggy code and are essential for pinpointing the buggy location when using SBFL approaches. 
As shown in Figure \ref{fig:stackTracesStructure}, stack traces have a pre-defined pattern. Therefore, we implement a regex-based parser to identify stack traces in bug reports. Our parser identifies the stack traces and extracts the file name, method name, and line number from each stack trace entry. 
Among all the 803 bugs we collected from Defects4J, we identified 89 bugs that have stack traces in the bug report. 

\begin{table}
\caption{An overview of our studied systems from Defects4J v2.0.0. {\em \#Total Bugs}, {\em \#Bugs with Stack Traces}, {\em \#Studied Bugs}, {\em LOC}, and {\em \#Tests} show the total number of bugs in Defects4J, the total number of bugs with stack traces, the number of bugs studied, lines of code, and tests in each system, respectively.}
\centering
\scalebox{1}{
\begin{tabular}{lrrrrr}
    \toprule
    \textbf{System} & \textbf{\#Total} & \textbf{\#Bugs with} & \textbf{\#Studied} & \textbf{LOC} & \textbf{\#Tests}\\
    &{\bf Bugs}&{\bf Stack Traces} & {\bf Bugs}\\
    \midrule
    Cli  & 39   & 3     & 2       & 4K     & 94     \\
    Closure & 174  & 8    & 8      & 90K    & 7,911   \\
    Codec & 18   & 1      & 1       & 7K     & 206    \\
    Collections & 4 & 1 & 1        & 65K    & 1,286 \\
    Compress & 47  & 11   & 11       & 9K     & 73    \\
    Csv & 16      & 1    & 2       &  2K    & 54     \\
    Gson & 18    & 3    & 3       & 14K    & 720    \\
    JacksonCore & 26 & 4  & 4       & 22K    & 206     \\
    JacksonDatabind & 112 & 27 & 10   & 4K  & 1,098     \\
    Jsoup  & 93   & 8    & 8       & 8K     & 139     \\
    JxPath & 22  & 1     & 1       & 25K    & 308      \\
    Lang & 64     & 4    & 4       & 22K    & 2,291   \\
    Math & 106    & 6     & 3      & 85K    & 4,378    \\
    Mockito & 38  & 8      & 2       & 11K    & 1,379   \\
    Time & 26     & 2     & 2       & 28K    & 4,041   \\
    \midrule
    \textbf{Total}& 803 & 89 & 60 & 380K & 24,302 \\
    \bottomrule
\end{tabular}}
\label{tab:dataset}
\vspace{-1em}
\end{table}

\subsubsection{Collecting Test Coverage}\label{data_collection:test_cov}

To obtain the test information for fault localization, we need to collect the test execution results, detailed test coverage, and the bug-fix patch. However, there are some limitations in Defects4J. Recent studies \citep{kabadi2023future, chen2023back} found that many bugs in Defects4J v1.0 contain tests from the ``future'' (i.e., added by developers after the bugs were fixed)  in their designated buggy commits. 
This is problematic because the test coverage may contain developers' knowledge of the bug, which can cause noise and bias in the result of the downstream research.
Because of this, we do not directly utilize Defects4J's data in our study. Instead, since we have the creation date of each bug report, we extract the commit right before the bug report creation from the project's repository. This commit represents the code that was available to developers when they started to address the bug report. We refer to this commit as a \textit{\textbf{bug report commit}} to differentiate it from the buggy commit provided by Defects4J.
\par
Due to the above-mentioned issues in Defects4J, our next step is to collect the test results and detailed test coverage for the bug report commits. To do that, we utilized GZoltar \citep{campos-etal-2012-gzoltar}. GZoltar is a debugging, fault localization, and coverage extraction tool for Java applications that was often applied to analyze Defects4J bugs \citep{silva-etal-2021-flacoco, 9609138, 7985698, 8918955}. For each crash report, we check out the bug report commit, compile the system, and execute the tests. 
We utilize the GZoltar CLI tool to implement a script to extract the test coverage and test execution results for the bugs from our study. Since there are many ancient commits, we encountered many challenges in compiling the system and running the tests. For example, since some commits in a project can use different versions of JVM, we had to implement a tool to automatically switch JVM versions when analyzing the commits. There are also many dependency-related issues, where a project uses different versions of a library across commits, but upgrading/downgrading the library version can cause dependency conflicts that lead to compilation errors.
In total, we spent over 100 hours compiling and collecting the data. Despite our best efforts, we excluded some bugs where we were not able to execute and collect the test coverage information. 
\par
Table \ref{tab:dataset} provides detailed information on the bugs studied in this work, i.e., bugs from Defect4J 2.0 with valid bug reports. In total, we include 803 bugs from 15 open-source projects. In this work, we target user-reported bugs with stack traces reported, i.e., crash-reporting bugs. 
Among the 15 projects, there are 89 crash-reporting bugs (i.e., bugs with stack traces in their bug reports), which represent 11.08\% of the total set. Of these 89 bugs, we were able to compile and collect the test execution information for 63 bugs. Finally, we excluded three bugs from our study whose root cause is not in a method, as this work focuses on method-level bug localization. The absence of buggy methods can happen in two different situations: (1) The bugfix involves altering lines outside the methods (for example, changing the value of a global variable); or (2) The bugfix includes the creation of new methods, but no existing method is updated. In total, we study \textbf{60 bugs}. Despite the low percentage of crash reports found in Defects4J, they match the percentage of bugs with stack traces in other projects \citep{chen-etal-2021-demystifying}. 
\subsection{Evaluation Metrics}\label{sec:metrics}
Previous studies suggest that performing fault localization at the file level lacks precision \citep{kochhar2016practitioners}, while opting for statement-level granularity can cause the approach to miss important code context \citep{parnin2011automated}. A survey of developers \citep{kochhar2016practitioners} further indicates that method-level granularity is the most preferred approach for fault localization tasks.  Hence, we conduct our analysis at the \textit{method-level} in this paper by following prior studies~\citep{b2016learning,autofl,benton2020effectiveness,lou2021boosting}.



\uhead{\em Mean Average Precision (MAP).}
MAP is a popular metric for evaluating the effectiveness of ranking systems, such as fault localization (FL) techniques~\citep{sohn2017fluccs,chen-etal-2021-pathidea,hirsch2023map}. It is the mean of the average precision of all faults. It reflects how well an FL technique ranks faulty (relevant) methods near the top. Precision at a given rank $k$ ($P@k$) is defined as the proportion of faulty methods among the top $k$ methods:
\[
P@k = \frac{\text{number of faulty methods in top } k \text{ ranks}}{k}
\]
For a given bug, the \emph{Average Precision (AP)} is computed as the average of the precision values at each rank where a faulty method appears:
\[
AP = \frac{1}{M} \sum_{k=1}^{N} P@k \cdot \text{rel}(k)
\]
where $M$ is the total number of faulty methods, $N$ is the total number of ranked methods, and $\text{rel}(k)$ is 1 if the method at rank $k$ is faulty, and 0 otherwise.
Finally, \emph{Mean Average Precision (MAP)} is obtained by averaging the AP values across all bugs in the dataset:
\[
MAP = \frac{1}{Q} \sum_{q=1}^{Q} AP_q
\]
where $Q$ is the total number of faults.
A higher MAP indicates that faulty methods are consistently ranked higher by the localization technique, reflecting better overall performance.



\uhead{\em Mean Reciprocal Rank (MRR).}
MRR is a ranking metric that evaluates how quickly the first faulty (relevant) method appears in the ranked list. For each fault $q$ (out of $Q$ total faults), let $r_q$ denote the rank position of its earliest (highest-ranked) faulty method. The reciprocal rank for that fault is then $1/r_q$.
Mean Reciprocal Rank (MRR) is calculated as the mean of these reciprocal ranks across all faults:
\[
\mathrm{MRR} = \frac{1}{Q} \sum_{q=1}^{Q} \frac{1}{r_q}
\]
A higher MRR indicates that, on average, faulty methods appear closer to the top of the ranking, reflecting better fault localization performance.

\uhead{\em Top K}. The Top K metric represents the number of bugs in which at least one buggy
method is located between the first K best-ranked methods in the approach. According to a previous study \citep{parnin2011automated}, developers only check a limited number of suspicious locations. Due to this, we use  1, 3, and 5 as the values for K.

\section{Results and Analysis}\label{sec:rqs}

This section discusses the results of our research questions (RQs). We present the motivation, approach, and results for each RQ.

\subsection*{RQ1 - Are the test failures related to the bug in crash reports?}
\label{sec:rq1}

\phead{Motivation.} 
Spectrum-Based Fault Localization (SBFL) relies on failing test coverage to identify faulty code, under the assumption that failing tests actually reveal the presence of a bug~\citep{abreu-etal-2007-accuracy, jones-harrold-2005-tarantula, naish2011model}. However, when a bug report is created, it is often unclear whether the tests that fail at that point are actually related to the bug described in the report-these are known as fault-triggering tests~\citep{just2014mutants, haben2023importance, chen2023back}. This question is especially important for crash-related bugs, where developers may only have the stack trace and available test results to guide debugging. With this in mind, in RQ1, we investigate whether the test failures observed at the bug report commit are actually caused by the bug described in the crash report. We then also explore how the presence or absence of such fault-triggering tests impacts the effectiveness of SBFL in localizing the fault. While analyzing fault-triggering tests that are later added to bug reports would reflect post-localization efforts, our goal in RQ1 is to focus on the reporting stage, characterizing the absence of such tests when the bug is first reported, and suggesting a practical solution for this early debugging scenario.

\phead{Approach.} 
To understand whether the test failures are related to the bugs in crash reports, we analyze the number of crash reports that contain failing fault-triggering tests. As discussed in Section~\ref{data_collection:test_cov}, we perform our study on the bug report commits for each bug report to capture the exact state of the project when the crash was first reported and to avoid biases of tests added after the bug was reported. Specifically, we execute all available tests using GZoltar to collect test execution results and coverage data, then compare the coverage of failing tests with the methods mentioned in the crash stack traces and the actual buggy methods identified in the fixes. Building on the prior work by \cite{chen2023back}, which revisited Defects4J to analyze fault-triggering tests and bug timelines, we further inspected the bug reports to determine whether they contain stack traces. Through this analysis, we identified 157 bugs without stack traces, allowing us to compare crash-report bugs with non-crash bugs.
\par
To study the impact of missing fault-triggering tests on SBFL techniques, we select Ochiai \citep{abreu-etal-2007-accuracy} as the baseline since it is widely used and has been shown to perform well on real faults \citep{7985698, 6676912, 4041886}. We utilize the detailed coverage and test results obtained through the Gzoltar execution to apply the Ochiai formula at the method level. The formula calculates the suspiciousness of each code statement, allowing us to rank the methods based on their likelihood of containing the bug. The SBFL suspiciousness formulas vary for each technique, but all are based on the idea that the code that is more covered by failing tests and less covered by passing tests is more likely to be buggy. 
The Ochiai formula assigns a suspiciousness score between 0 (not suspect) and 1 (highly suspect) for each piece of code element (e.g., class, method, or statement). If applied at a method level, we calculate the suspicious score of a given method \textit{j} as:
\begin{equation}
\label{eq:ochiai}
s_o = \frac{n_{11}(j)}{\sqrt{(n_{11}(j) + n_{01}(j))*(n_{11}(j) + n_{10}(j))}}
\end{equation}

\begin{table}[t]
    \centering
    \caption{\label{tab:Table1} Description of the terms in the Ochiai formula. \textit{Covered} indicates if the component (in our case, the method) was executed or not during the testing (i.e., is covered or not by the test). \textit{Test} indicates if the test case failed or passed during its execution.}
    \large
    \begin{tabular}{*{3}{c}}
        \hline
         \textbf{n} & \textbf{Covered} & \textbf{Test} \\
        \hline
       $n_{00}$ & no & passed \\
        \hline 
        $n_{10}$ & yes & passed \\
        \hline
         $n_{01}$ & no & failed \\
        \hline
         $n_{11}$ & yes & failed \\
        \hline
    \end{tabular}
\end{table}

\noindent in which the terms are defined in Table ~\ref{tab:Table1}. 
Each term \textit{n} in the formula (e.g., $n_{00}$) corresponds to the number of tests that obey specific criteria for the method \textit{j} for which the suspiciousness score is being calculated.
The first criterion determines the method coverage, whether the method is covered by a test ($n_{1x}$) or not ($n_{0x}$). The second criterion indicates the execution status, passing ($n_{x0}$) or failing ( $n_{x1}$).
So, if a test covers the method \textit{j} and its execution fails, for instance, this test will be computed under $n_{11}$.
As a calculation example, suppose that we have the following test case results for a given method M1:
\begin{itemize}
    \item 6 failing test cases that cover M1 ($n_{11}$)
    \item 2 failing test cases that do not cover M1 ($n_{01}$)
    \item 10 passing test cases that cover M1 ($n_{10}$)
\end{itemize}
By applying Equation \ref{eq:ochiai}, we get that the Ochiai score for method M1 is approximately 0.530.

\phead{Results.}
\textbf{Only 3.33\% of the crash report bugs include failing tests that are directly caused by the bug (i.e., fault-triggering tests).}
Table \ref{tab:fault-triggering-tests} shows the test execution and fault-triggering test results for the crash report bugs. When comparing with the bugs without stack traces, we obtain the following numbers:
\begin{itemize}
    \item \textbf{Crash Reports:} 2 out of 60 bugs have fault-triggering tests \textbf{(3.33\%)}
    \item \textbf{Bugs without Stack Traces:} 16 out of 157 bugs have fault-triggering tests \textbf{10.19\%)}
\end{itemize}
\par

\begin{table}
\caption{Test execution results for the crash report bugs. {\em \#Bugs}, {\em \#Tests} and {\em \#Bugs with Fault-triggering Tests} show the total number of studied bugs, the total amount of tests and the total number of bugs with fault-triggering tests; while  {\em Failing Tests (Avg.)}, and {\em Fault-triggering Tests (Avg.)} represent the average number of failing tests and fault-triggering tests, respectively, calculated among the bugs.}
\centering
\scalebox{0.85}{
\begin{tabular}{lrrrr}
    \toprule
    \textbf{System (\#Bugs) } & \textbf{\#Tests} & \textbf{Failing Tests} & \textbf{Fault-triggering} & {\bf \#Bugs with Fault-}\\
     & & {\bf (Avg.)} & {\bf Tests (Avg.)} & {\bf Triggering Tests} \\
    \midrule
    Cli (2)  & 94   &  2    &    0   &    0   \\
    Closure (8) & 7,911  &  2.125   &    0  &    0   \\
    Codec (1) & 206   &  0     &    0 &    0   \\
    Compress (11) & 73  & 1  &   0   &    0   \\
    Csv (1) & 54     &  0   &    0  &    0  \\
    Gson (3) & 720    &  0   &   0  &    0   \\
    JacksonCore (4) & 206 & 0.5  &   0.25  &    1  \\
    JacksonDatabind (10) & 1,098 & 26.9 &   0.1  &    1  \\
    Jsoup (8)  & 139   &  0.125   &    0   &    0   \\
    JxPath (1) & 308  &  0    &      0   &    0 \\
    Lang (4) & 2,291     &  8.5   &    0   &    0 \\
    Math (3) & 4,378    &  0.667    &   0  &    0 \\
    Mockito (2) &  1,379  &  21.5     &   0  &    0\\
    Time (2) & 4,041     &   20.5   &     0  &    0 \\
    \midrule
    \textbf{Total (60)}& 24,302 & 5.987 & 0.025 & 2  \\
    \bottomrule
\end{tabular}}
\label{tab:fault-triggering-tests}
\vspace{-1em}
\end{table}

Based on the findings, we observed that most crash report bugs (96.67\%) either lack fault-triggering tests or do not trigger them (i.e., fault-triggering tests did not fail). In addition, we can see that the percentage of fault-triggering tests in crash report bugs is significantly lower (3.33\%) compared to the set of bugs without stack traces (10.19\%). The reason may be that such exception-related issues are more often triggered during production, rather than during testing. Another reason may be that developers are more likely to handle exceptions for debugging purposes rather than fault prevention \citep{shah2008developers}. This shift in exception handling as a debugging aid increases the likelihood of exception-related issues occurring in production. 
In such cases, applying traditional SBFL techniques would not be ideal, given the tool's inability to differentiate between buggy and non-buggy statements when no test failures cover them. 
Due to the absence of failing fault-triggering tests, the runtime information included in the bug report (i.e., stack traces) is the final resource for assistance. In short, studying how to leverage stack traces for fault localization is an essential supplement to traditional SBFL techniques that use failing fault-triggering tests.


\begin{tcolorbox}
Only 3.33\% of crash-report bugs have failing tests that expose the bug  (i.e., fault-triggering tests), limiting the applicability of coverage-based fault localization.
\end{tcolorbox}

To study how the absence of fault-triggering tests impacts SBFL, we apply Ochiai to our studied bugs. To evaluate the approach, we utilize the metrics presented in Section \ref{sec:metrics}, specifically Top-1, Top-3, Top-5, MAP, and MRR. We exclude bugs without failing tests from the Ochiai results, given that the presence of failing tests is necessary to compute the suspiciousness score. Out of 60 bugs, 23 did not have any failing tests. 
Note that even if a test fails, it may not be related to the bug we are interested in (i.e., it is not a fault-triggering test).

\noindent\textbf{On the bug report commit, SBFL performed poorly in all the projects, only being able to locate 2/60 bugs among the Top-5 methods.}
Table \ref{tab:rq1-ochiai} shows the SBFL results applied to the bug report commit for the crash report bugs that contain failing tests. Ochiai could not locate any bug in Top-1 and located only 1 in Top-3 and 2 in Top-5. These two localized bugs are from the \textit{JacksonCore} and \textit{Compress} projects, with \textit{JacksonCore} containing a fault-triggering test. The highest MAP and MRR values across projects are 0.0693 and 0.0836, respectively, in the JacksonCore project. Considering that the values of MAP and MRR range between 0 and 1, the obtained values are extremely low. 
In contrast, a prior study by~\citet{chen-etal-2022-useful} that evaluates Ochiai on the original Defects4J benchmark achieved an average MAP of 0.30 and an average MRR of 0.42, which is significantly higher than our results obtained when there is no failing fault-triggering tests.

The MAP and MRR values assess the effectiveness of fault localization techniques in returning relevant results in the top ranking. These metrics are essential indicators that demonstrate the usefulness of fault localization techniques. According to a prior survey by~\citet{kochhar2016practitioners}, 80\% of developers consider a fault localization technique successful if it can localize bugs in the top 5 positions. Hence, our finding shows the inefficiency and limitations of SBFL techniques due to the lack of failing tests related to the bug reports. 
\begin{table} 
    \caption{Ochiai results for all the studied systems.{\em \#Bugs} represents the total number of studied bugs, while {\em \#Bugs with FT} is the number of bugs in which there is at least one Failing Test.}
    \centering
    \label{tab:rq1-ochiai}
    \scalebox{0.8}{\setlength{\tabcolsep}{0.4cm}
    \begin{tabular}{l|rrrrr}
    \toprule
        \textbf{System (\#Bugs | \#Bugs with FT) } & \textbf{Top-1} & \textbf{Top-3} & \textbf{Top-5} & \textbf{MAP} & \textbf{MRR} \\ \midrule
       
        \textbf{Cli} (2 | 2) & 0  & 0 & 0 & 0.0045 & 0.0045  \\ 
        \midrule
        \textbf{Closure} (8 | 8) & 0  & 0 & 0 & 0.0007 & 0.0012 \\  
        \midrule
        \textbf{Codec} (1 | 0) & -  & - & - & - & - \\  
        \midrule
        \textbf{Collections} (0 | 0) & -  & - & -  & - & - \\  
        \midrule
        \textbf{Compress} (11 | 6) & 0 & 0 & 1 & 0.0079 & 0.023 \\  
        \midrule
        \textbf{Csv} (1 | 0) & -  & - & - & - & - \\  
        \midrule
        \textbf{Gson} (3 | 0) & -  & - & -  & - & - \\  
        \midrule
        \textbf{JacksonCore} (4 | 2) & 0 & 1 & 1 & 0.0693 & 0.0836 \\  
        \midrule
        \textbf{JacksonDatabind} (10 | 8) & 0  & 0 & 0 & 0.0005 & 0.0005 \\  
        \midrule
        \textbf{Jsoup} (8 | 1) & 0  & 0 & 0 & 0 & 0 \\  
        \midrule
        \textbf{JxPath} (1 | 0) & -  & - & -  & - & - \\  
        \midrule
        \textbf{Lang} (4 | 4) & 0  & 0 & 0 & 0.0008 & 0.0008 \\  
        \midrule
        \textbf{Math} (3 | 2) & 0  & 0 & 0 & 0.0002 & 0.0001 \\  
        \midrule
        \textbf{Mockito} (2 | 2) & 0  & 0 & 0 & 0.0060 & 0.0281 \\  
        \midrule
        \textbf{Time} (2 | 2) & 0  & 0 & 0 & 0.0003 & 0.0003 \\          
        \bottomrule
    \end{tabular}}
\end{table}

\begin{tcolorbox}
Due to a lack of fault-triggering tests, traditional SBFL techniques have inferior localization results. 
\end{tcolorbox}

\subsection*{RQ2 - What is the relationship between stack traces and buggy location? }

\phead{Motivation.}
In the previous RQ, we observed that most crash report bugs lack fault-triggering tests. As a result, stack traces are the only information available to reflect the code execution at the time of failure. In this RQ, we investigate the relationship between the stack traces included in bug reports and the actual buggy code. Specifically, we aim to assess whether stack traces provide meaningful guidance for identifying the faulty location and understanding the nature of the fix. To do this, we analyze two dimensions: (i) the type of modification made in the fix since this reflects how developers interpreted and responded to the exception, and (ii) the proximity between the modified (buggy) methods and the stack trace methods. These dimensions allow us to evaluate the extent to which stack traces directly support fault localization and resolution.

\begin{figure*}
    \centering
\includegraphics[scale=0.4]{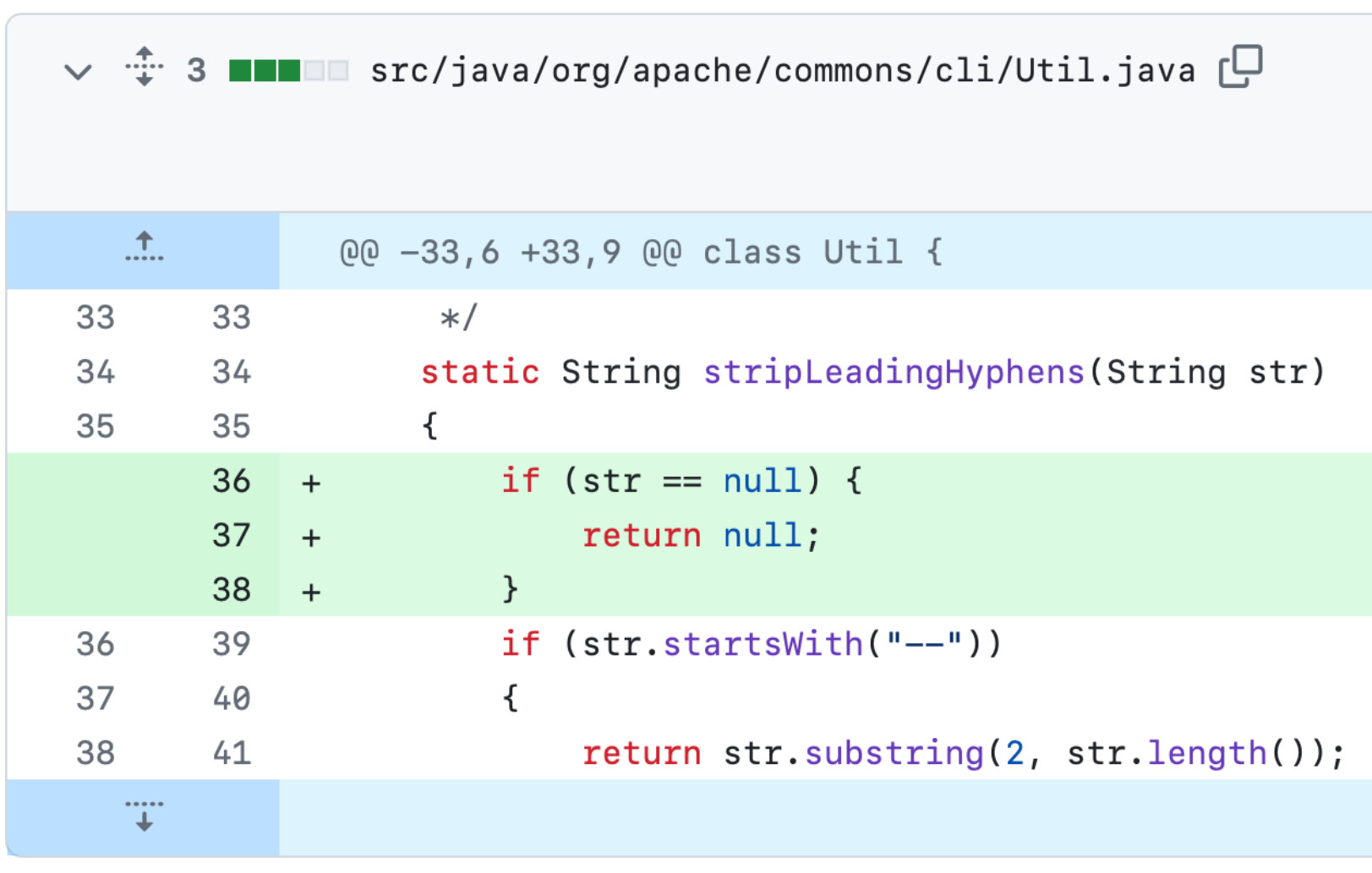}
    \caption{Extract from the bugfix patch for Cli-5, classified in the Exception Prevention category.}
    \label{fig:exception-prevention-example}
\end{figure*}

\begin{figure*}
    \centering
\includegraphics[scale=0.4]{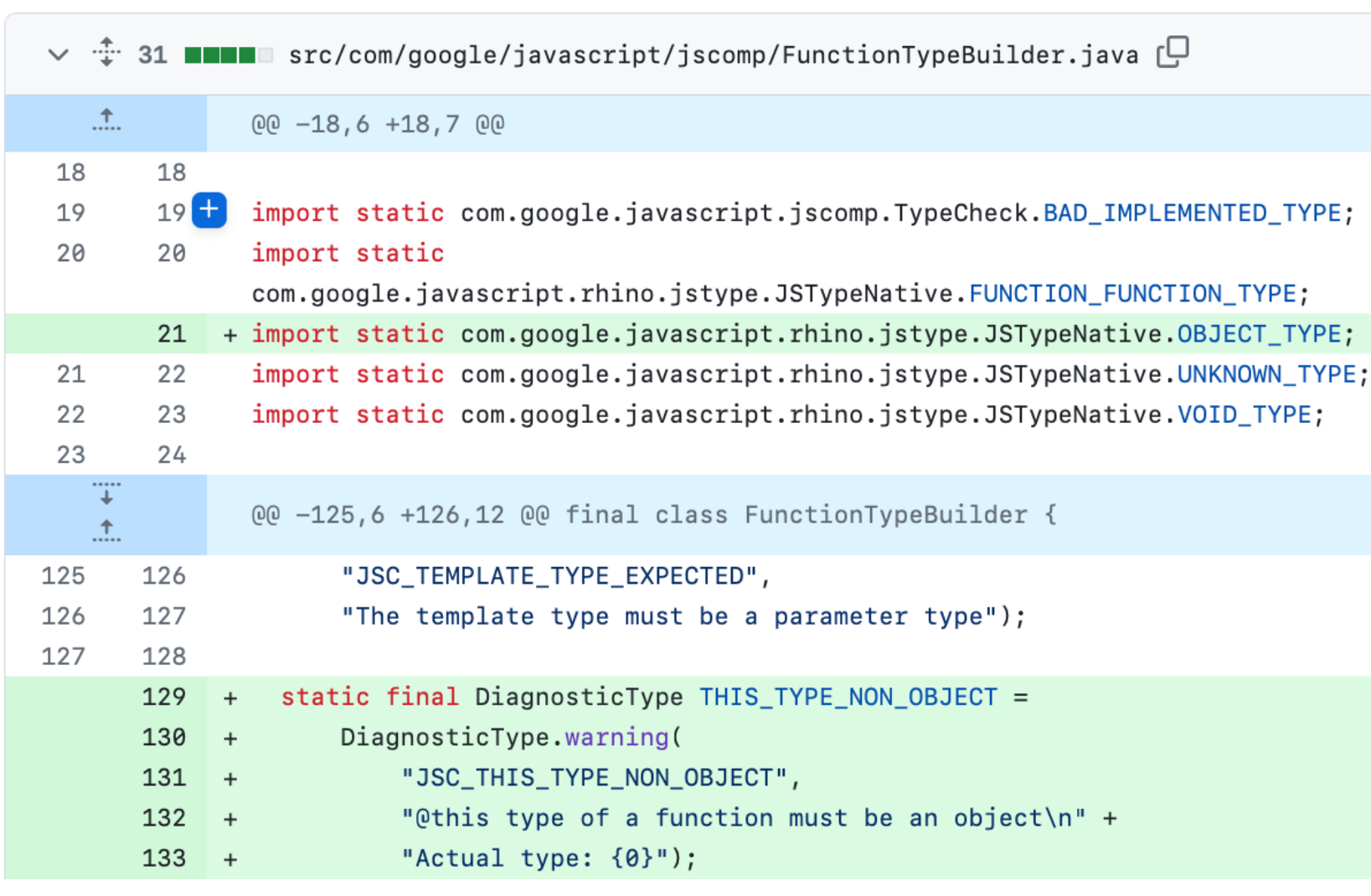}
    \caption{Extract from the bugfix patch for Closure-152, classified in the Exception Conversion category.}
    \label{fig:exception-conversion-example}
\end{figure*}

\begin{figure*}
    \centering
\includegraphics[scale=0.4]{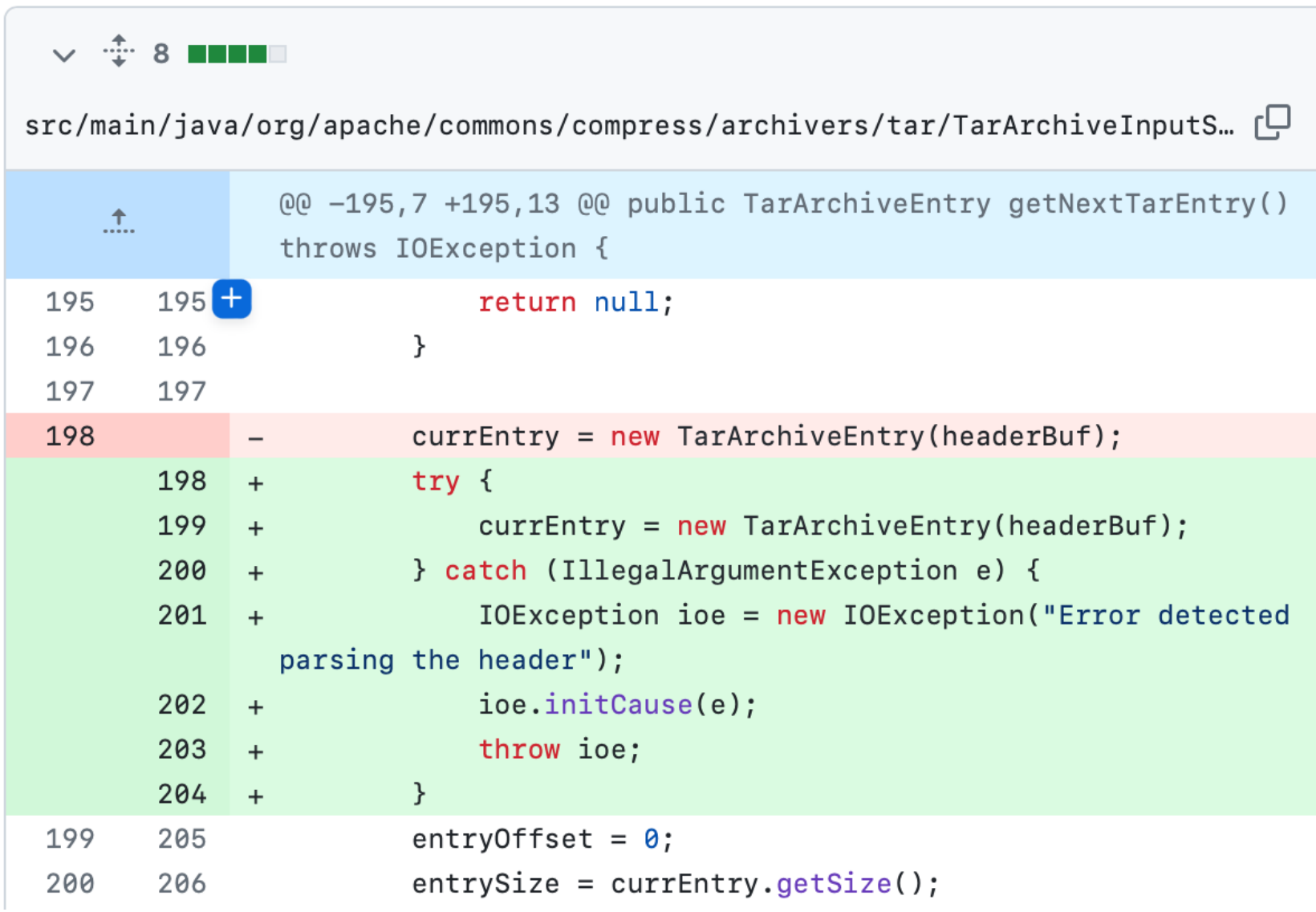}
    \caption{Extract from the bugfix patch for COMPRESS-12, classified in the Exception Wrapping category.}
    \label{fig:exception-wrapping-example}
\end{figure*}

\phead{Approach.}
To study the type of modification performed in the bug fix to handle the exception, we manually examine the bug-fix patch of all the studied bugs. Based on the modification performed, we classify the bug-fix intention type into four categories:
\begin{enumerate}
    \item \textbf{Exception Prevention:} This category includes code modifications to prevent a specific exception's recurrence. For example, the bugfix patch from \textit{Cli-5} introduced a new conditional structure to avoid the occurrence of the reported \textit{NullPointerException} (Figure \ref{fig:exception-prevention-example}).
    \item \textbf{Exception Conversion:} In this type of bugfix, the exception is converted into a warning or error message. For instance, to fix the bug \textit{Closure-152} (Figure \ref{fig:exception-conversion-example}), the developers handled the \textit{ClassCastException} being thrown and created a warning detailed message.
    \item \textbf{Exception Wrapping:} This category represents the cases in which the developers fix the bug by wrapping the exception into another exception type. One example is the bug \textit{COMPRESS-12}, in which the exception \textit{IllegalArgumentException} was wrapped into the exception \textit{IOException} (Figure \ref{fig:exception-wrapping-example})
    \item \textbf{Keep Throwing:} Finally, this category includes the bugs in which no exception handling-related code is found in the bugfix. 
\end{enumerate}

\begin{figure*}
    \centering
\includegraphics[scale=0.45]{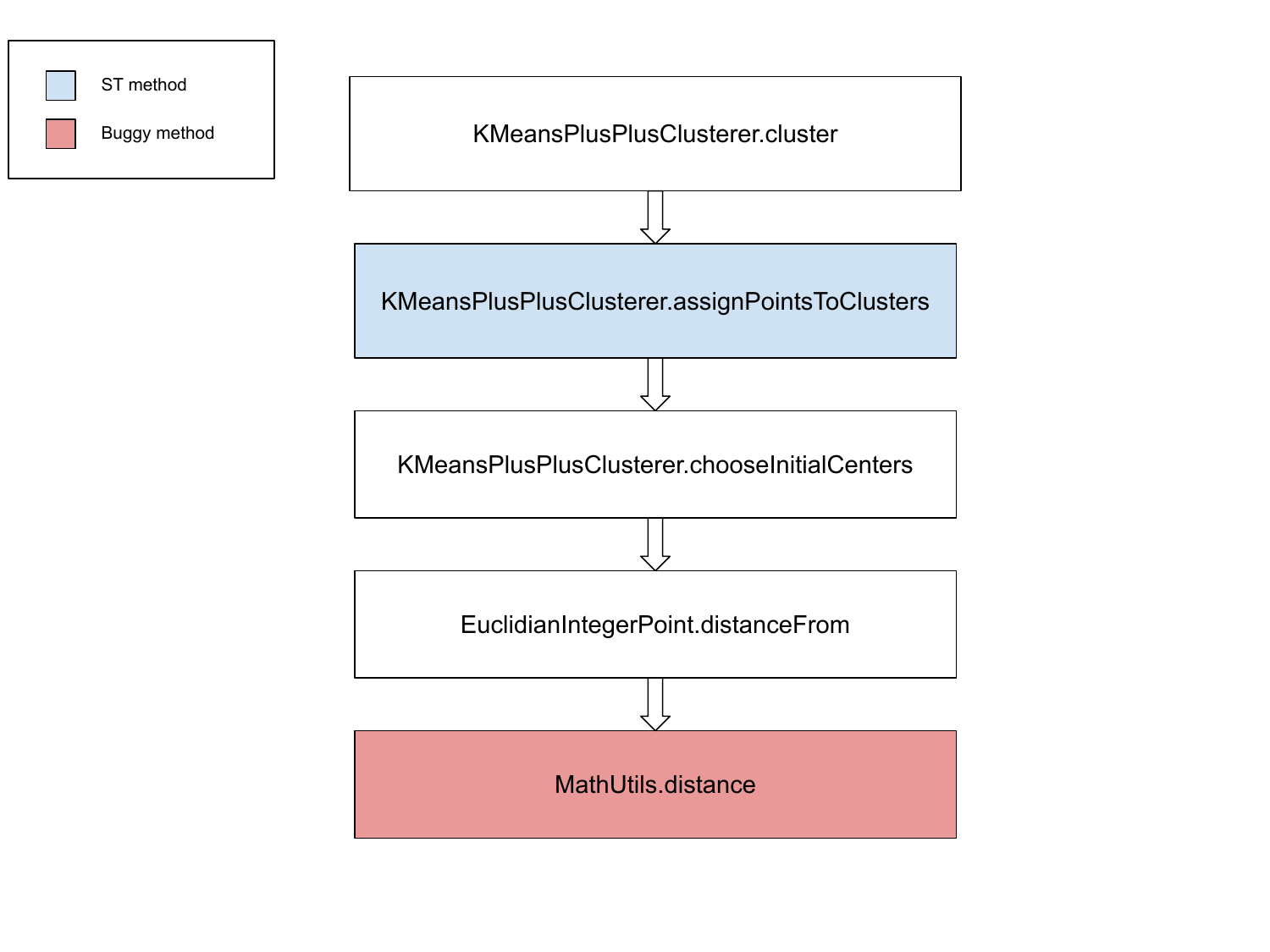}
    \caption{Call Graph for the bug Math-79, which has a 3-call distance between the stack trace and the buggy method.}
    \label{fig:call-graph-example}
\end{figure*}

To study the distance between the stack trace and the buggy methods, we utilize the stack trace content and the code in the bug report commit to create a call graph for each bug. We then measure the minimum distance between a method in the stack trace and one of the buggy methods, if reachable. For example, the bug Math-79 in Figure \ref{fig:call-graph-example} has a 3-call distance between the stack trace and the buggy method. The distance is zero for a bug if the buggy method is recorded in the stack trace. 

\phead{Results.}  \textbf{In 83\% of cases, the bug fix intention is to prevent the exception from happening again.} 
In the majority of the fixes (83.3\%), the intention behind the fix is \textbf{Exception Prevention} (i.e., to prevent the exception from happening again). This means that the stack traces are directly related to the root cause of the bug, suggesting that the stack trace is an essential source of information for locating the bug. The following two predominant categories are \textbf{Exception Wrapping} (10.0\%) and \textbf{Exception Conversion} (5.0\%), which also possess a correlation between the fix and the stack traces. In only one bug, \textit{COMPRESS-31}, the exception-related code was not touched for the fix (\textbf{Keep Throwing} category). After a more detailed investigation, we noticed that, despite the bug being in the Defect4j database, its resolution field was marked as `Not A Problem` on Jira, which explains why no action was taken regarding the exception.

\noindent\textbf{In 78.3\%  of the bugs, the buggy methods are directly reachable from the stack trace, with an average distance of 0.34 method calls}. 
Upon examining the proximity between stack traces and buggy methods, our analysis reveals that 66.67\% of the bugs feature at least one buggy method directly present in the stack trace, indicating a zero distance. Furthermore, a total of 78.3\% of the bugs contain buggy methods that are reachable from the stack trace (i.e., distance is zero or more), averaging a very short distance from the stack trace methods of 0.34 method calls. In such scenarios, the short distance between the stack trace and buggy methods greatly reduces the search space for developers or automated tools. This makes stack traces particularly effective for fault localization, as they often point to or near the actual faulty method, even in the absence of explicit failing tests. This observation supports our proposed SBEST approach, which uses stack trace methods to approximate fault-triggering behavior. 
We conduct a manual analysis of the remaining 21.7\% of the bugs for which the buggy methods are unreachable from the stack trace. In these cases, we find that the stack traces only contain external entries (i.e., entries referring to external libraries) or the provided stack traces were incomplete. In summary, our findings underscore the potential of utilizing stack traces for fault localization. These findings indicate that stack-trace proximity is not only common but also practically useful as a substitute for inadequate test coverage in spectrum-based fault localization when such tests are absent.

\begin{tcolorbox}
In 98.3\% of the studied bugs, the developer’s fix was applied to one of the methods shown in the exception’s stack trace (Exception Prevention, Exception Conversion, or Exception Wrapping). Additionally, 78.3\% of the buggy methods are reachable from the stack traces, with an average distance of 0.34 method calls. This shows that the stack traces are a valuable source of information about the bug.
\end{tcolorbox}

\subsection*{RQ3 - Can we utilize the stack traces to help detect the buggy locations?}
\label{sec:rq3}
\par
\phead{Motivation.}
In RQ1, we found that the existing failing tests rarely matched the fault-triggering tests. Without fault-triggering tests, the performance of SBFL techniques is greatly affected. In addition, in RQ2, we found that the stack traces are a
valuable source of information about the bug. The intention behind the bugfix is usually related to the exception, and the buggy methods are often a short distance away from the stack traces. Prior studies \citep{chen-etal-2021-demystifying, chen-etal-2021-pathidea} show that developers usually rely on stack traces when investigating bugs, as they provide essential information about the buggy location. Stack traces, similar to fault-triggering tests, carry contextual and execution information associated with the root causes of bugs. In a way, stack traces can represent the coverage of a failing fault-triggering test. 
Therefore, in this RQ, we aim to investigate how stack traces can be leveraged to complement test cases in locating bugs. We conduct this analysis on the same set of 60 crash-reporting bugs identified earlier.
 
\phead{Approach.} 
To understand how stack trace information can contribute to locating buggy methods, we propose to evaluate an approach called SBEST (Spectrum-Based localization Enhanced by Stack Traces). SBEST is a fault localization approach based on SBFL, incorporating stack trace information with the test coverage data. This technique applies SBFL principles, but instead of using the failing fault-triggering test coverage in the Ochiai formula, it considers the methods that appear on the stack trace entries as the source of the failure. We define a process causing a test failure if it appears on any of the frames in a stack trace. 
Prior studies \citep{wong2014boosting, chen-etal-2021-pathidea} have also shown that the position of methods in the stack trace can be helpful in fault localization. Therefore, our approach incorporates a Stack Trace (ST) score. 

We compare our approach with four baselines. The first is Ochiai, a widely adopted and high-performing SBFL technique~\citep{7985698}. The second is the stack trace position baseline, where methods appearing higher in the stack trace are ranked as more suspicious. The third is Call Frequency-Based Fault Localization (CFBFL)~\citep{callfreq}, which replaces binary hit-based coverage with the frequency of method appearances in unique call stack contexts. CFBFL leverages execution information to capture deeper relationships between stack traces and faulty behavior, making it suitable even when explicit failing tests are unavailable or weak. To fairly incorporate this method into our evaluation, we first examined the program versions that immediately precede the bug report, consistent with our SBEST setup. We then implemented CFBFL using its best-performing variant, JaccardC, as reported in their original study~\citep{callfreq}. Lastly, we evaluate the SB\_score, a partial version of our approach that incorporates coverage from stack-trace-related methods but excludes stack trace position weighting.

\noindent Our overall approach, SBEST, consists of the sum of two scores: the SB score and the ST score.

\uhead{Spectrum Based (SB) Score.}
The SB score is based on an approach designed to utilize information from stack traces in the fault localization process. It is based on the Ochiai formula and utilizes existing test coverage information. However, instead of using the failing tests as a hint of the bug location (since there may not be any test failure or the failure is not related to the reported bug, as found in RQ1), we use the tests covering the methods in the stack trace as the fault-triggering tests. While we use Ochiai in this study due to its strong performance and wide adoption, our approach is not tied to this specific formula; it is general and can be applied using any SBFL formula.

Our key insight is to use the stack trace as a proxy for failing test behavior by identifying tests that cover the methods appearing in the stack trace. These tests are then treated as fault-triggering inputs, enabling the application of SBFL even in the absence of actual failing tests. While we use the Ochiai formula in our evaluation due to its strong performance and popularity, our approach is not tied to any specific formula. The technique is general and can be applied with any SBFL formula, as the novelty lies in how we construct coverage-like information from stack traces rather than in the formula itself. 

We base this change on the fact that the stack trace represents the system’s execution at the moment of failure, in the same way that the fault-triggering tests do. More specifically, we utilize the methods present in the first five internal entries from the stack traces to select the proxy failing tests that will be applied to Ochiai’s formula. We base this design on the fact that the higher the stack trace entry, the more probable it is to be related to the bug~\citep{wong2014boosting}.

The selection of proxy failing tests is determined by counting the number of lines each existing test covers in the Stack Trace methods. We first compute the number of covered lines from these methods for each test by applying the following formula:

\begin{equation}
\label{eq:st_covered_lines}
ST\_Covered\_lines\_number_t = \sum_{m=1}^{5} CL_m
\end{equation}

\noindent in which \textit{m} represents each of the top 5 methods in the Stack Trace, and $CL_{m}$ is the number of lines in the corresponding method covered by the test. 
Then, we select the set of 
\textit{X} highest results by applying the following formula, in which T represents the set of tests and X is a threshold for the number of selected tests:
\begin{equation}
\label{eq:fake_failing_tests}
\begin{split}
T_{\text{failing}} = \{ t \in T \mid t \text{ is one of the } X \text{ tests with the highest } \\ ST\_Covered\_lines\_number_t \}
\end{split}
\end{equation}

\noindent We noticed that selecting a value for \textit{X} that was too low often missed buggy locations because it was too focused on just a few methods. On the other hand, a high threshold also generates inaccurate results since the approach cannot differentiate between the methods in the stack trace. Experimenting showed that setting it to 15 gave the best results. Finally, we apply the Ochiai formula (\ref{eq:ochiai}) using these 15 tests as failing tests and all the remaining as passing tests. Table \ref{tab:failing_tests_results} presents the experimental results for different values of proxy failing tests \textit{X}, which will be discussed further in the results section. 

\par
\uhead{Stack Trace (ST) Score.}
The SB score described above merges the information in the stack traces with the test coverage to compute the suspiciousness score for a given method. In addition, we use the ST score used in a previous study \citep{chen-etal-2021-pathidea} to boost the stack trace's impact on the calculated score. To do this, after having computed the SB\_score following the approach described before, we add to it a ST\_score calculated in the following manner:
\begin{equation}
\label{eq:st_score}
ST\_score = 
\begin{cases} 
\frac{1}{ST\_rank} & \text{if } ST\_rank \leq 10 \\
0.1 & \text{if } ST\_rank > 10 \\
0 & \text{if } \text{method not found}
\end{cases}
\end{equation}

\noindent In which \textit{ST\_rank} is the position where the given method appears in the stack trace after the external entries are removed. For example, if the method appears in the second position, the ST\_score would be 1/2=0.5. 
\par
\uhead{Suspiciousness Score.}
After having both the SB Ochiai score and the ST Score, we compute the suspiciousness score of a given method by applying the formula:
\begin{equation}
\label{eq:suspcioness_score}
Suspiciousness\_score = SB\_Ochiai\_score + ST\_score
\end{equation}

\par
\uhead{Suspiciousness Rank.} After having the final suspiciousness score, we generate the final suspiciousness rankings. The higher the suspiciousness score, the higher the ranking. For instance, if the highest calculated suspiciousness score is 1.0 for the method \textit{M8}, \textit{M8} will be the first method in the suspiciousness rank, as it has the highest probability of containing the bug.

\begin{table} 
    \caption{Fault localization results for all the studied projects. {\em \#Bugs} represents the total number of studied bugs in each project.}
    \centering
    \label{tab:rq3-techniques}
    \vspace{-3mm}
    \scalebox{0.68}{\setlength{\tabcolsep}{0.6cm}
    \begin{tabular}{l|l|rrrrr}
    \toprule
        \textbf{System (\#Bugs) }& \textbf{Technique} & \textbf{Top-1} & \textbf{Top-3} & \textbf{Top-5} & \textbf{MAP} & \textbf{MRR} \\
        \midrule
        
        \textbf{Cli} (2) & Ochiai & 0  & 0 & 0 & 0.005 & 0.005  \\
        \textbf{} & Stack Trace & 1 & 1 & 2  & 0.6 & 0.6 \\ 
         \textbf{} & SB\_score & 0 & 0 & 1 & 0.133 & 0.133 \\ 
         \textbf{} & CFBFL & 0 & 0 & 1 & 0.222 & 0.222 \\ 
          \textbf{} & SBEST  & 1 & 1 & 2  & 0.6 & 0.6 \\ 
          
        \midrule
        \textbf{Closure} (8) & Ochiai  & 0  & 0 & 0 & 0.001 & 0.001 \\  
        \textbf{} & Stack Trace & 1 & 2 & 2  & 0.171 & 0.217 \\ 
         \textbf{} & SB\_score & 0 & 0 & 0 & 0.0003 & 0.005 \\ 
         \textbf{} & CFBFL & 0 & 0 & 1 & 0.006 & 0.0056 \\ 
          \textbf{} & SBEST  & 1 & 2 & 2  & 0.146 & 0.167 \\     
        
        \midrule
        \textbf{Codec} (1) & Ochiai & 0  & 0 & 0 & - & - \\  
        \textbf{} & Stack Trace & 1 & 1 & 1  & 0.333 & 1 \\ 
         \textbf{} & SB\_score & 1 & 1 & 1 & 0.389 & 1 \\ 
         \textbf{} & CFBFL & 1 & 1 & 1 & 0.399 & 1 \\ 
          \textbf{} & SBEST  & 1 & 1 & 1  & 0.585 & 1 \\ 
          
        \midrule
        \textbf{Compress} (11) & Ochiai & 0 & 0 & 1 & 0.008 & 0.023 \\  
        \textbf{} & Stack Trace & 3 & 6 & 8  & 0.412 & 0.443 \\ 
         \textbf{} & SB\_score & 0 & 3 & 3 & 0.141 & 0.159 \\ 
         \textbf{} & CFBFL & 0 & 1 & 1 & 0.002 & 0.002 \\ 
          \textbf{} & SBEST  & 3 & 9 & 9  & 0.462 & 0.491 \\
          
        \midrule
        \textbf{Csv} (1) & Ochiai & 0  & 0 & 0 & - & - \\  
        \textbf{} & Stack Trace & 0 & 1 & 1  & 0.333 & 0.333 \\ 
         \textbf{} & SB\_score & 1 & 1 & 1 & 1 & 1 \\ 
         \textbf{} & CFBFL & 0 & 1 & 1 & 0.014 & 0.019 \\ 
          \textbf{} & SBEST  & 1 & 1 & 1 & 1 & 1 \\
          
        \midrule
        \textbf{Gson} (3) & Ochiai & 0  & 0 & 0  & - & - \\  
        \textbf{} & Stack Trace & 1 & 2 & 2  & 0.52 & 0.52 \\ 
         \textbf{} & SB\_score & 1 & 1 & 1 & 0.334 & 0.334 \\ 
         \textbf{} & CFBFL & 0 & 0 & 0 & 0.0001 & 0.0001 \\ 
          \textbf{} & SBEST  & 1 & 2 & 2 & 0.51 & 0.51 \\
          
        \midrule
        \textbf{JacksonCore} (4) & Ochiai & 0 & 1 & 1 & 0.069 & 0.084 \\  
        \textbf{} & Stack Trace & 1 & 1 & 1  & 0.25 & 0.25 \\ 
         \textbf{} & SB\_score & 0 & 0 & 0 & 0.041 & 0.046 \\ 
         \textbf{} & CFBFL & 0 & 0 & 1 & 0.001 & 0.001 \\ 
          \textbf{} & SBEST  & 0 & 1 & 1 & 0.161 & 0.164 \\
          
        \midrule
        \textbf{JacksonDatabind} (10) & Ochiai & 0  & 0 & 0 & 0.001 & 0.001 \\  
        \textbf{} & Stack Trace & 1 & 2 & 3  & 0.169 & 0.245 \\ 
         \textbf{} & SB\_score & 0 & 1 & 1 & 0.013 & 0.021 \\
         \textbf{} & CFBFL & 1 & 1 & 1 & 0.176 & 0.215 \\ 
          \textbf{} & SBEST  & 1 & 3 & 3 & 0.036 & 0.078 \\
          
        \midrule
        \textbf{Jsoup} (8) & Ochiai & 0  & 0 & 0 & 0 & 0 \\   
        \textbf{} & Stack Trace & 3 & 3 & 5  & 0.441 & 0.445 \\ 
         \textbf{} & SB\_score & 0 & 0 & 0 & 0.022 & 0.032 \\
         \textbf{} & CFBFL & 2 & 3 & 4 & 0.224 & 0.299 \\ 
          \textbf{} & SBEST  & 3 & 4 & 4 & 0.407 & 0.428 \\
          
        \midrule
        \textbf{JxPath} (1)  & Ochiai & 0  & 0 & 0  & - & - \\   
        \textbf{} & Stack Trace & 1 & 1 & 1  & 0.333 & 0.333 \\ 
         \textbf{} & SB\_score & 0 & 0 & 0 & 0.063 & 0.063 \\ 
         \textbf{} & CFBFL & 0 & 0 & 0 & 0.001 & 0.001 \\ 
          \textbf{} & SBEST   & 1 & 1 & 1 & 1 & 1 \\
          
        \midrule
        \textbf{Lang} (4) & Ochiai & 0  & 0 & 0 & 0.001 & 0.001 \\  
        \textbf{} & Stack Trace & 1 & 3 & 3  & 0.438 & 0.438 \\ 
         \textbf{} & SB\_score & 0 & 1 & 1 & 0.141 & 0.141 \\ 
         \textbf{} & CFBFL & 0 & 0 & 1 & 0.01 & 0.01 \\ 
          \textbf{} & SBEST  & 2 & 3 & 3 & 0.584 & 0.584 \\
          
        \midrule
        \textbf{Math} (3)  & Ochiai & 0  & 0 & 0 & 0.0002 & 0.0001 \\  
        \textbf{} & Stack Trace & 2 & 2 & 2  & 0.222 & 0.458 \\ 
         \textbf{} & SB\_score & 0 & 0 & 0 & 0.041 & 0.029 \\ 
         \textbf{} & CFBFL & 0 & 1 & 1 & 0.113 & 0.129 \\ 
          \textbf{} & SBEST  & 2 & 2 & 2 & 0.137 & 0.347 \\
          
        \midrule
        \textbf{Mockito} (2) & Ochiai & 0  & 0 & 0 & 0.006 & 0.028 \\  
        \textbf{} & Stack Trace & 0 & 1 & 1  & 0.024 & 0.167 \\ 
         \textbf{} & SB\_score & 0 & 1 & 1 & 0.037 & 0.25 \\ 
         \textbf{} & CFBFL & 0 & 0 & 0 & 0.002 & 0.002 \\ 
          \textbf{} & SBEST  & 0 & 1 & 1 & 0.037 & 0.25 \\
          
        \midrule
        \textbf{Time} (2) & Ochiai & 0  & 0 & 0 & 0.0003 & 0.0003 \\     
        \textbf{} & Stack Trace & 0 & 1 & 2  & 0.292 & 0.292 \\ 
         \textbf{} & SB\_score & 1 & 1 & 1 & 0.5 & 0.5 \\ 
         \textbf{} & CFBFL & 0 & 0 & 0 & 0.001 & 0.001 \\ 
          \textbf{} & SBEST  & 0 & 1 & 1 & 0.333 & 0.333 \\ 
          
        \midrule
        
        \textbf{Total} (60) & Ochiai & 0  & 1 & 2 & 0.009  & 0.014 \\ 
        \textbf{} & Stack Trace & 16 & 27 & 34  & 0.324 & 0.41 \\ 
         \textbf{} & SB\_score & 4 & 10 & 11 & 0.204 & 0.265 \\ 
         \textbf{} & CFBFL & 4 & 8 & 13 & 0.084 & 0.136 \\ 
          \textbf{} & SBEST  & 17 & 32 & 33 & 0.429 & 0.497 \\ 
          \bottomrule
    \end{tabular}}
\end{table}

\begin{table}[ht]
\centering
\caption{Fault Localization Performance with Different Numbers of Failing Tests for SBEST}
\label{tab:failing_tests_results}
\scalebox{0.8}{\setlength{\tabcolsep}{0.6cm}
\begin{tabular}{c|cccccc}
\hline
\textbf{X (Proxy Failing Tests)} & \textbf{Top-1} & \textbf{Top-3} & \textbf{Top-5} & \textbf{MAP} & \textbf{MRR} \\
\hline
5  & 14 & 22 & 27 & 0.37111 & 0.43887 \\
10 & 16 & 29 & 30 & 0.39121 & 0.43944 \\
\textbf{15} & \textbf{17} & \textbf{32} & \textbf{33} & \textbf{0.42846} & \textbf{0.49647} \\
20 & 14 & 23 & 28 & 0.37532 & 0.42598 \\
25 & 12 & 20 & 25 & 0.30897 & 0.41720 \\
\hline
\end{tabular}
}
\end{table}

\begin{table}[ht]
\centering
\caption{Fault Localization Performance for Different Numbers of Top Stack-Trace Methods for SBEST}
\label{tab:sbest_stackmethods_results}
\scalebox{0.8}{\setlength{\tabcolsep}{0.5cm}
\begin{tabular}{c|cccccc}
\hline
\textbf{Top \#Stack-trace Methods} & \textbf{Top-1} & \textbf{Top-3} & \textbf{Top-5} & \textbf{MAP} & \textbf{MRR} \\
\hline
\textbf{5}  & \textbf{17} & \textbf{32} & \textbf{33} & \textbf{0.42846} & \textbf{0.49647} \\
10 & 7  & 10 & 11 & 0.38962 & 0.41266 \\
15 & 4  & 7  & 9  & 0.20011 & 0.2409 \\
\hline
\end{tabular}
}
\end{table}

\phead{Results.} \textbf{The Stack Traces ranking alone locates 34 out of the 60 bugs in the Top-5.} Table \ref{tab:rq3-techniques} presents the metrics results for the 2 baselines as well as for the SB\_score alone and the SBEST approach for all the systems. We can see that the worst-performing technique is Ochiai, which, as discussed before, locates only 2 bugs in the Top 5. \textbf{The best-performing approach is SBEST, but the Stack Traces ranking is a close second.} The Stack Trace ranking locates 16 bugs in Top-1, 27 in Top-3, and 34 in Top-5. Surprisingly, the Stack Traces alone provide very good results, locating more than half of the bugs (56.67\%) within the Top 5. This shows that stack traces are a significant source of information about bug location and should be prioritized when studying FL in crash reports. The SB\_score can locate some bugs that the Stack Trace ranking does not locate, but in general, it does not perform very well, identifying only 18.33\% of the buggy methods in the Top 5.

CFBFL, which ranks methods based on their frequency in unique stack contexts during execution, identified 4 bugs in Top-1, 8 in Top-3, and 13 in Top-5. While it improves over Ochiai, it lags behind Stack Trace ranking and SBEST in all metrics. The MAP and MRR of CFBFL (0.084 and 0.136) are significantly lower than those of SBEST and Stack Traces, suggesting that while call frequency provides useful contextual signals, it may not be sufficient on its own for localizing crash faults when fault-triggering test coverage is sparse or missing.


We observe that adding the ST score improves the SB\_score. The resulting technique, SBEST, outperforms the Stack Traces approach in the Top 1 and Top 3 ranks, locating 17 and 32 bugs (compared to Stack Traces’ 16 and 27), and only one fewer bug in the Top 5, for a total of 33. Overall, SBEST and Stack Traces are the two most effective localization techniques across all projects. Stack Traces performs better on JacksonCore, while SBEST achieves the best results on Lang. On average, SBEST attains a MAP of 0.42846 and an MRR of 0.49647, showing moderate improvements (32.22\% and 17.43\%) over the Stack Trace baseline in scenarios where fault-triggering tests are unavailable. These results indicate that stack traces can effectively complement passing test information in identifying root causes.

\begin{tcolorbox}
Overall, out of the 60 studied bugs, SBEST successfully located 17 bugs in the Top 1, 32 in the Top 3, and 33 in the Top 5. Additionally, it achieves improvements of 32.22\% on MAP and 17.43\% on MRR compared to the Stack-Trace-based ranking.
\end{tcolorbox}

\textbf{SBEST performs best with 15 proxy failing tests and the top 5 stack trace methods.} Tables~\ref{tab:failing_tests_results} and~\ref{tab:sbest_stackmethods_results} present the fault localization performance of SBEST under different parameter settings for the number of proxy failing tests and the number of stack trace methods, respectively. FL performance is stable between 10-20 proxy failing tests, with Top-1 results peaking at 17 faults. However, the performance degrades when we include more than 15 proxy failing tests or more than 5 stack trace methods. This suggests that, when too many failing tests or stack-trace methods are included, more noise may be introduced, which can negatively impact fault localization accuracy. For example, using 25 proxy failing tests or 15 stack methods significantly reduces Top-1 and MAP scores.

\begin{tcolorbox}
    Setting the number of proxy failing tests (X) to 15 and using the top 5 stack trace methods gives the best results, with 17 bugs in Top-1 and a MAP of 0.4285. Other configurations show a drop in performance.
\end{tcolorbox}

\section{Discussion and Implications of Our Findings}
\label{sec:discussion}
\phead{On the use of stack traces in the absence of failing tests.} From the results obtained in Section \ref{sec:rqs}, we observe that most crash report bugs do not contain fault-triggering tests. In such cases, the test failure information is unavailable for the fault localization process, causing techniques such as SBFL to fail. Conversely, the stack traces are deeply correlated with the cause of the bug. We found that, in 66\% of the bugs, at least one buggy method is directly listed in the stack trace. Even when this is not the case, the buggy methods are usually just a few calls away from it.  In addition, the most common bugfix intention is to prevent the exception in the stack trace from happening again, highlighting the significant association between them.
\par
\phead{Leveraging stack trace rankings for enhanced fault localization.} In the real world, test failures are not always present. Especially regarding production-phase bugs, it is essential to look for alternative sources of information for fault localization. Using the data available in logs and stack traces is a good path for such scenarios. Although previous studies have utilized stack traces for Fault Localization, many are based on information retrieval (IRFL) approaches \citep{zhou2012should, saha2013improving, lam2017bug}. IRFL techniques treat all content in the bug report, including stack traces, as textual information; therefore, they miss important context, such as the ranking of the stack traces. Our results show that the stack traces ranking alone could locate more than half of the bugs within Top-5 and, thus, should be better utilized.
\par
\phead{Encouraging the use of stack traces to complement execution information.} SBEST, our approach combining the stack trace ranking with the code coverage, shows promising results. The improvement in the metrics implies that the coverage information enhanced the stack trace ranking, aiding in obtaining a better overview of the system execution at the moment of failure.  Our primary goal in studying this technique was to gain a deeper understanding of the scenario surrounding these bugs and how each piece of available information can aid in fault localization. We believe this is just the first step, and that future research can benefit from the findings above to develop more sophisticated techniques.  In addition, we make all the data from this study available at Zenodo\footnote{\url{https://zenodo.org/records/11062413}}, including the detailed code coverage and test results of the bug report commits that took us 100 hours to obtain. We believe this data will be beneficial for future research on fault localization, automatic bug fixing, test generation, and other related areas. Our data excludes any post-fix developer knowledge, which would allow researchers to tailor tools more closely aligned with real-world scenarios.
\par
\phead{Limitations.} While our findings highlight the potential of stack traces for fault localization in the absence of fault-triggering tests, we acknowledge several limitations. Our study focuses specifically on bugs that result in exceptions with available stack traces. In practice, many bugs do not manifest as crashes or produce exceptions, which may limit the applicability of our findings to the broader space of software faults. Future work should investigate the applicability of stack-trace-based localization to a wider range of bug types and software systems.

\phead{Future Directions.} 
Our findings open several avenues for future research. One particularly important direction is to explore how execution signals such as stack traces and runtime logs can be better leveraged in the absence of fault-triggering tests. These lightweight signals can serve as a proxy for fault-triggering execution, allowing fault localization methods like SBFL to operate in real-world scenarios where fault-triggering tests are missing. Additionally, future studies could expand evaluation beyond Defects4J to improve generalizability across projects and programming languages. Some recent research, such as AutoFL~\citep{autofl}, applies LLMs to reason over code and coverage data for fault localization. At the same time, LIBRO~\citep{libro} generates bug-reproducing tests directly from natural language bug reports with stack traces. Building on these ideas, one direction is to leverage LLMs to interpret the underlying causes of failures from stack traces and coverage data, and use that information to filter out irrelevant methods or code regions. This could enable more accurate and efficient ranking of suspicious methods, potentially supporting automated patch generation and thereby enhancing the practicality of SBEST in scenarios where explicit test failures are absent.

\section{Threats to Validity}\label{sec:threat}
In this section, we discuss the threats to validity related to this study.

\subsection{External Validity.} 

Threats to external validity relate to the generalizability of our findings. To minimize this threat, we selected 15 systems from Defects4J, as used by ~\cite{just-etal-2014-defects4j}, a widely used benchmark of real bugs in prior studies. These systems vary in size, test coverage, and domain. However, we acknowledge that our study focuses on 60 crash-report bugs ($\approx$10\% of Defects4J) with available stack traces. This proportion aligns with observations from real-world studies, which show that only a limited subset of user-reported bugs contain stack traces~\citep{bettenburg2008makes,chen-etal-2021-demystifying}. Although this represents a subset of all bugs, crash reports are among the most practically significant cases developers face, as stack traces often provide the first diagnostic evidence when failures occur and have been widely utilized in fault localization research \citep{bettenburg2008makes,zou2021flfamily}. We prioritized crash-reporting bugs for which stack traces and test coverage could be reliably extracted, though this was sometimes challenging due to legacy and dependency issues. We publicly release our dataset to support replication and further study. Finally, while our evaluation focuses on Java projects, the proposed approach can be extended to other programming languages in future work.

\subsection{ Internal validity.} 
Threats to internal validity relate to the extent to which the results of the study can be attributed solely to the experimental treatments and not to flaws in the experimental design. In this study, we only analyze Defects4J bugs. Although Defects4J is widely used \citep{lutellier-etal-2020-coconut, chen-etal-2021-pathidea, li2021fault, ye2022neural}, the findings might be different in other systems, especially non-Java systems.
Another threat to internal validity is how we selected the bug reports with stack traces. We collect stack traces in the description and comments, which is the typical place in which they are added. However, in some cases, developers may also add stack traces via attachments. Nonetheless, this is very rare, representing less than 1\% of the cases based on a previous study from \citet{chen-etal-2021-demystifying}.

\subsection{Construct validity.} 
Construct validity refers to how well the study's procedures and metrics accurately capture the concepts they intend to investigate.
We use three evaluation metrics in our study: Top K, MAP, and MRR. These metrics are commonly used in FL and have been used in many previous studies \citep{8873606, chen-etal-2021-pathidea, xia2016automated, kochhar2016practitioners}.

\section{Conclusion}\label{sec:conclusion}
Spectrum-based fault Localization (SBFL) approaches are highly recognized due to their accuracy, efficiency, and simplicity. However, their effectiveness depends on the availability of fault-triggering tests. In the absence of such test failures, alternative approaches should be considered. In this paper, we investigate the use of stack traces for SBFL in the absence of fault-triggering tests. We study 60 crash reports from the Defects4J benchmark. We find that most of the studied crash reports do not contain fault-triggering tests. This results in very low efficiency when using traditional SBFL. On the other hand, when it comes to the stack traces, we find that in 98.3\% of the studied bugs, the bugfix intention is directly correlated with the exception in the stack trace. In addition, in 78.3\% of the bugs, the buggy method is directly reachable from the stack trace, with an average distance of only 0.34 method calls. This indicates that buggy methods are typically very close to the stack traces in terms of the execution call graph. Furthermore, our results demonstrate that even without employing advanced techniques, the stack traces alone provide a reliable indication of buggy locations, capable of locating more than half of the bugs in the Top 5. Finally, we develop a simplified SBFL method called SBEST that utilizes stack trace information as a proxy for failing tests, thereby integrating it with coverage information to perform SBFL. 

SBEST located 17 bugs in Top-1, 32 in Top-3, and 33 in Top-5, with MAP and MRR values that are 32.22\% and 17.43\% higher than the Stack Trace ranking, respectively. The results suggest that combining coverage information with stack traces can enhance fault localization by offering a clearer view of system execution when fault-triggering tests are unavailable. Future studies can benefit from the findings of this research to develop more intricate techniques. In addition, all the data that we have made available can help studies in areas such as automatic bug fixing, test generation, and, of course, fault localization. As potential future work, we plan to expand our analysis by increasing our bug dataset by utilizing other sources beyond Defects4J.


\section{Declarations}
\section*{Conflicts of Interest.}
All authors are from Concordia University and the University of Alberta. There may be conflicts with individuals from these two institutions.

\section*{Funding.}
The authors have no funding to report.

\section*{Data Availability Statements.} We make the data publicly available online for replication and encourage future research in this area \citep{sbest}.



\bibliographystyle{natbib}
\bibliography{paper}

\begin{thebibliography}{}

\bibitem[Abreu {\em et~al.}(2006)Abreu, Zoeteweij, and Van~Gemund]{4041886}
Abreu, R., Zoeteweij, P., and Van~Gemund, A.~J. (2006).
\newblock An evaluation of similarity coefficients for software fault localization.
\newblock In {\em 2006 12th Pacific Rim International Symposium on Dependable Computing (PRDC'06)\/}, pages 39--46.

\bibitem[Abreu {\em et~al.}(2007)Abreu, Zoeteweij, and van Gemund]{abreu-etal-2007-accuracy}
Abreu, R., Zoeteweij, P., and van Gemund, A.~J. (2007).
\newblock On the accuracy of spectrum-based fault localization.
\newblock In {\em Testing: Academic and industrial conference practice and research techniques-MUTATION (TAICPART-MUTATION 2007)\/}, pages 89--98. IEEE.

\bibitem[Abreu {\em et~al.}(2009)Abreu, Zoeteweij, and Van~Gemund]{abreu2009spectrum}
Abreu, R., Zoeteweij, P., and Van~Gemund, A.~J. (2009).
\newblock Spectrum-based multiple fault localization.
\newblock In {\em 2009 IEEE/ACM International Conference on Automated Software Engineering\/}, pages 88--99. IEEE.

\bibitem[B.~Le {\em et~al.}(2016)B.~Le, Lo, Le~Goues, and Grunske]{b2016learning}
B.~Le, T.-D., Lo, D., Le~Goues, C., and Grunske, L. (2016).
\newblock A learning-to-rank based fault localization approach using likely invariants.
\newblock In {\em Proceedings of the 25th international symposium on software testing and analysis\/}, pages 177--188.

\bibitem[Benton {\em et~al.}(2020)Benton, Li, Lou, and Zhang]{benton2020effectiveness}
Benton, S., Li, X., Lou, Y., and Zhang, L. (2020).
\newblock On the effectiveness of unified debugging: An extensive study on 16 program repair systems.
\newblock In {\em Proceedings of the 35th IEEE/ACM International Conference on Automated Software Engineering\/}, pages 907--918.

\bibitem[Beszedes {\em et~al.}(2020)Beszedes, Horváth, Di~Penta, and Gyimothy]{levcontextfl}
Beszedes, A., Horváth, F., Di~Penta, M., and Gyimothy, T. (2020).
\newblock Leveraging contextual information from function call chains to improve fault localization.
\newblock In {\em 2020 IEEE 27th International Conference on Software Analysis, Evolution and Reengineering (SANER)\/}, pages 468--479.

\bibitem[Bettenburg {\em et~al.}(2008)Bettenburg, Just, Schr{\"o}ter, Weiss, Premraj, and Zimmermann]{bettenburg2008makes}
Bettenburg, N., Just, S., Schr{\"o}ter, A., Weiss, C., Premraj, R., and Zimmermann, T. (2008).
\newblock What makes a good bug report?
\newblock In {\em Proceedings of the 16th ACM SIGSOFT International Symposium on Foundations of software engineering\/}, pages 308--318.

\bibitem[Campos {\em et~al.}(2012)Campos {\em et~al.}]{campos-etal-2012-gzoltar}
Campos, J. {\em et~al.} (2012).
\newblock Gzoltar: an eclipse plug-in for testing and debugging.
\newblock In {\em Proceedings of the 27th IEEE/ACM international conference on automated software engineering\/}, pages 378--381.

\bibitem[Chen {\em et~al.}(2021a)Chen, Chen, and Wang]{chen-etal-2021-demystifying}
Chen, A.~R., Chen, T.-H., and Wang, S. (2021a).
\newblock Demystifying the challenges and benefits of analyzing user-reported logs in bug reports.
\newblock {\em Empirical Software Engineering\/}, {\bf 26}, 1--30.

\bibitem[Chen {\em et~al.}(2021b)Chen, Chen, and Wang]{chen-etal-2021-pathidea}
Chen, A.~R., Chen, T.-H., and Wang, S. (2021b).
\newblock Pathidea: Improving information retrieval-based bug localization by re-constructing execution paths using logs.
\newblock {\em IEEE Transactions on Software Engineering\/}, {\bf 48}(8), 2905--2919.

\bibitem[Chen {\em et~al.}(2022)Chen, Chen, and Chen]{chen-etal-2022-useful}
Chen, A.~R., Chen, T.-H., and Chen, J. (2022).
\newblock How useful is code change information for fault localization in continuous integration?
\newblock In {\em 37th IEEE/ACM International Conference on Automated Software Engineering\/}, pages 1--12.

\bibitem[Cui {\em et~al.}(2020)Cui, Jia, Chen, Zheng, and Liu]{9201269}
Cui, Z., Jia, M., Chen, X., Zheng, L., and Liu, X. (2020).
\newblock Improving software fault localization by combining spectrum and mutation.
\newblock {\em IEEE Access\/}, {\bf 8}, 172296--172307.

\bibitem[de~Oliveira {\em et~al.}(2018)de~Oliveira, Camilo-Junior, de~Andrade~Freitas, and Vincenzi]{de2018ftmes}
de~Oliveira, A. A.~L., Camilo-Junior, C.~G., de~Andrade~Freitas, E.~N., and Vincenzi, A. M.~R. (2018).
\newblock Ftmes: A failed-test-oriented mutant execution strategy for mutation-based fault localization.
\newblock In {\em 2018 IEEE 29th International Symposium on Software Reliability Engineering (ISSRE)\/}, pages 155--165. IEEE.

\bibitem[de~Souza {\em et~al.}(2016)de~Souza, Chaim, and Kon]{de2016spectrum}
de~Souza, H.~A., Chaim, M.~L., and Kon, F. (2016).
\newblock Spectrum-based software fault localization: A survey of techniques, advances, and challenges.
\newblock {\em arXiv preprint arXiv:1607.04347\/}.

\bibitem[Gong {\em et~al.}(2012)Gong, Lo, Jiang, and Zhang]{gong2012interactive}
Gong, L., Lo, D., Jiang, L., and Zhang, H. (2012).
\newblock Interactive fault localization leveraging simple user feedback.
\newblock In {\em 2012 28th IEEE International Conference on Software Maintenance (ICSM)\/}, pages 67--76. IEEE.

\bibitem[Gong {\em et~al.}(2014)Gong, Zhang, Seo, and Kim]{gong2014locating}
Gong, L., Zhang, H., Seo, H., and Kim, S. (2014).
\newblock Locating crashing faults based on crash stack traces.
\newblock {\em arXiv preprint arXiv:1404.4100\/}.

\bibitem[Haben {\em et~al.}(2023)Haben, Habchi, Papadakis, Cordy, and Traon]{haben2023importance}
Haben, G., Habchi, S., Papadakis, M., Cordy, M., and Traon, Y.~L. (2023).
\newblock The importance of discerning flaky from fault-triggering test failures: A case study on the chromium ci.
\newblock {\em arXiv preprint arXiv:2302.10594\/}.

\bibitem[Hirsch and Hofer(2023)Hirsch and Hofer]{hirsch2023map}
Hirsch, T. and Hofer, B. (2023).
\newblock The map metric in information retrieval fault localization.
\newblock In {\em 2023 38th IEEE/ACM International Conference on Automated Software Engineering (ASE)\/}, pages 1480--1491. IEEE.

\bibitem[Hong {\em et~al.}(2017)Hong, Kwak, Lee, Jeon, Ko, Kim, and Kim]{hong2017museum}
Hong, S., Kwak, T., Lee, B., Jeon, Y., Ko, B., Kim, Y., and Kim, M. (2017).
\newblock Museum: Debugging real-world multilingual programs using mutation analysis.
\newblock {\em Information and Software Technology\/}, {\bf 82}, 80--95.

\bibitem[Jiang {\em et~al.}(2012)Jiang, Li, Li, Zhang, Zhang, and Liu]{jiang2012stack}
Jiang, S., Li, W., Li, H., Zhang, Y., Zhang, H., and Liu, Y. (2012).
\newblock Fault localization for null pointer exception based on stack trace and program slicing.
\newblock In {\em 2012 12th International Conference on Quality Software\/}, pages 9--12.

\bibitem[Jones and Harrold(2005)Jones and Harrold]{jones-harrold-2005-tarantula}
Jones, J.~A. and Harrold, M.~J. (2005).
\newblock Empirical evaluation of the tarantula automatic fault-localization technique.
\newblock In {\em Proceedings of the 20th IEEE/ACM international Conference on Automated software engineering\/}, pages 273--282.

\bibitem[Just {\em et~al.}(2014a)Just, Jalali, Inozemtseva, Ernst, Holmes, and Fraser]{just2014mutants}
Just, R., Jalali, D., Inozemtseva, L., Ernst, M.~D., Holmes, R., and Fraser, G. (2014a).
\newblock Are mutants a valid substitute for real faults in software testing?
\newblock In {\em Proceedings of the 22nd ACM SIGSOFT International Symposium on Foundations of Software Engineering\/}, pages 654--665.

\bibitem[Just {\em et~al.}(2014b)Just, Jalali, and Ernst]{just-etal-2014-defects4j}
Just, R., Jalali, D., and Ernst, M.~D. (2014b).
\newblock Defects4j: A database of existing faults to enable controlled testing studies for java programs.
\newblock In {\em Proceedings of the 2014 international symposium on software testing and analysis\/}, pages 437--440.

\bibitem[Kabadi {\em et~al.}(2023)Kabadi, Kong, Xie, Bao, Prana, Le, Le, and Lo]{kabadi2023future}
Kabadi, V., Kong, D., Xie, S., Bao, L., Prana, G. A.~A., Le, T.-D.~B., Le, X.-B.~D., and Lo, D. (2023).
\newblock The future can’t help fix the past: Assessing program repair in the wild.
\newblock In {\em 2023 IEEE International Conference on Software Maintenance and Evolution (ICSME)\/}, pages 50--61. IEEE.

\bibitem[Kang {\em et~al.}(2024a)Kang, Yoon, Askarbekkyzy, and Yoo]{libro}
Kang, S., Yoon, J., Askarbekkyzy, N., and Yoo, S. (2024a).
\newblock Evaluating diverse large language models for automatic and general bug reproduction.
\newblock {\em IEEE Trans. Softw. Eng.}, {\bf 50}(10), 2677–2694.

\bibitem[Kang {\em et~al.}(2024b)Kang, An, and Yoo]{autofl}
Kang, S., An, G., and Yoo, S. (2024b).
\newblock A quantitative and qualitative evaluation of llm-based explainable fault localization.
\newblock {\em Proc. ACM Softw. Eng.}, {\bf 1}(FSE).

\bibitem[Kochhar {\em et~al.}(2016)Kochhar, Xia, Lo, and Li]{kochhar2016practitioners}
Kochhar, P.~S., Xia, X., Lo, D., and Li, S. (2016).
\newblock Practitioners' expectations on automated fault localization.
\newblock In {\em Proceedings of the 25th international symposium on software testing and analysis\/}, pages 165--176.

\bibitem[Kuma {\em et~al.}(2020)Kuma, Higo, Matsumoto, and Kusumoto]{kuma2020improving}
Kuma, T., Higo, Y., Matsumoto, S., and Kusumoto, S. (2020).
\newblock Improving the accuracy of spectrum-based fault localization for automated program repair.
\newblock In {\em Proceedings of the 28th International Conference on Program Comprehension\/}, pages 376--380.

\bibitem[Küçük {\em et~al.}(2019)Küçük, Henderson, and Podgurski]{8918955}
Küçük, Y., Henderson, T. A.~D., and Podgurski, A. (2019).
\newblock The impact of rare failures on statistical fault localization: The case of the defects4j suite.
\newblock In {\em 2019 IEEE International Conference on Software Maintenance and Evolution (ICSME)\/}, pages 24--28.

\bibitem[Lam {\em et~al.}(2017)Lam, Nguyen, Nguyen, and Nguyen]{lam2017bug}
Lam, A.~N., Nguyen, A.~T., Nguyen, H.~A., and Nguyen, T.~N. (2017).
\newblock Bug localization with combination of deep learning and information retrieval.
\newblock In {\em 2017 IEEE/ACM 25th International Conference on Program Comprehension (ICPC)\/}, pages 218--229. IEEE.

\bibitem[Le {\em et~al.}(2013)Le, Thung, and Lo]{6676912}
Le, T.-D.~B., Thung, F., and Lo, D. (2013).
\newblock Theory and practice, do they match? a case with spectrum-based fault localization.
\newblock In {\em 2013 IEEE International Conference on Software Maintenance\/}, pages 380--383.

\bibitem[Le~Goues {\em et~al.}(2012)Le~Goues, Dewey-Vogt, Forrest, and Weimer]{le2012systematic}
Le~Goues, C., Dewey-Vogt, M., Forrest, S., and Weimer, W. (2012).
\newblock A systematic study of automated program repair: Fixing 55 out of 105 bugs for \$8 each.
\newblock In {\em 2012 34th International Conference on Software Engineering (ICSE)\/}, pages 3--13.

\bibitem[Li {\em et~al.}(2019)Li, Li, Zhang, and Zhang]{li2019deepfl}
Li, X., Li, W., Zhang, Y., and Zhang, L. (2019).
\newblock Deepfl: Integrating multiple fault diagnosis dimensions for deep fault localization.
\newblock In {\em Proceedings of the 28th ACM SIGSOFT international symposium on software testing and analysis\/}, pages 169--180.

\bibitem[Li {\em et~al.}(2021)Li, Wang, and Nguyen]{li2021fault}
Li, Y., Wang, S., and Nguyen, T.~N. (2021).
\newblock Fault localization with code coverage representation learning.
\newblock In {\em Proceedings of the 43rd International Conference on Software Engineering\/}, ICSE '21, page 661–673. IEEE Press.

\bibitem[Li {\em et~al.}(2007)Li, Harman, and Hierons]{li2007search}
Li, Z., Harman, M., and Hierons, R.~M. (2007).
\newblock Search algorithms for regression test case prioritization.
\newblock {\em IEEE Transactions on software engineering\/}, {\bf 33}(4), 225--237.

\bibitem[Liu {\em et~al.}(2018)Liu, Li, Zhao, and Gong]{liu2018optimal}
Liu, Y., Li, Z., Zhao, R., and Gong, P. (2018).
\newblock An optimal mutation execution strategy for cost reduction of mutation-based fault localization.
\newblock {\em Information Sciences\/}, {\bf 422}, 572--596.

\bibitem[Lou {\em et~al.}(2020)Lou, Ghanbari, Li, Zhang, Zhang, Hao, and Zhang]{lou2020can}
Lou, Y., Ghanbari, A., Li, X., Zhang, L., Zhang, H., Hao, D., and Zhang, L. (2020).
\newblock Can automated program repair refine fault localization? a unified debugging approach.
\newblock In {\em Proceedings of the 29th ACM SIGSOFT International Symposium on Software Testing and Analysis\/}, pages 75--87.

\bibitem[Lou {\em et~al.}(2021)Lou, Zhu, Dong, Li, Sun, Hao, Zhang, and Zhang]{lou2021boosting}
Lou, Y., Zhu, Q., Dong, J., Li, X., Sun, Z., Hao, D., Zhang, L., and Zhang, L. (2021).
\newblock Boosting coverage-based fault localization via graph-based representation learning.
\newblock In {\em Proceedings of the 29th ACM Joint Meeting on European Software Engineering Conference and Symposium on the Foundations of Software Engineering\/}, pages 664--676.

\bibitem[Lutellier {\em et~al.}(2020)Lutellier {\em et~al.}]{lutellier-etal-2020-coconut}
Lutellier, T. {\em et~al.} (2020).
\newblock Coconut: combining context-aware neural translation models using ensemble for program repair.
\newblock In {\em Proceedings of the 29th ACM SIGSOFT international symposium on software testing and analysis\/}, pages 101--114.

\bibitem[Moon {\em et~al.}(2014)Moon, Kim, Kim, and Yoo]{moon2014ask}
Moon, S., Kim, Y., Kim, M., and Yoo, S. (2014).
\newblock Ask the mutants: Mutating faulty programs for fault localization.
\newblock In {\em 2014 IEEE Seventh International Conference on Software Testing, Verification and Validation\/}, pages 153--162. IEEE.

\bibitem[Motwani and Brun(2023)Motwani and Brun]{motwani2023better}
Motwani, M. and Brun, Y. (2023).
\newblock Better automatic program repair by using bug reports and tests together.
\newblock In {\em 2023 IEEE/ACM 45th International Conference on Software Engineering (ICSE)\/}, pages 1225--1237. IEEE.

\bibitem[Naish {\em et~al.}(2011)Naish, Lee, and Ramamohanarao]{naish2011model}
Naish, L., Lee, H.~J., and Ramamohanarao, K. (2011).
\newblock A model for spectra-based software diagnosis.
\newblock {\em ACM Transactions on software engineering and methodology (TOSEM)\/}, {\bf 20}(3), 1--32.

\bibitem[Papadakis and Le~Traon(2012)Papadakis and Le~Traon]{papadakis2012using}
Papadakis, M. and Le~Traon, Y. (2012).
\newblock Using mutants to locate" unknown" faults.
\newblock In {\em 2012 IEEE Fifth International Conference on Software Testing, Verification and Validation\/}, pages 691--700. IEEE.

\bibitem[Papadakis and Le~Traon(2015)Papadakis and Le~Traon]{papadakis2015metallaxis}
Papadakis, M. and Le~Traon, Y. (2015).
\newblock Metallaxis-fl: mutation-based fault localization.
\newblock {\em Software Testing, Verification and Reliability\/}, {\bf 25}(5-7), 605--628.

\bibitem[Parnin and Orso(2011)Parnin and Orso]{parnin2011automated}
Parnin, C. and Orso, A. (2011).
\newblock Are automated debugging techniques actually helping programmers?
\newblock In {\em Proceedings of the 2011 international symposium on software testing and analysis\/}, pages 199--209.

\bibitem[Pearson {\em et~al.}(2017)Pearson, Campos, Just, Fraser, Abreu, Ernst, Pang, and Keller]{7985698}
Pearson, S., Campos, J., Just, R., Fraser, G., Abreu, R., Ernst, M.~D., Pang, D., and Keller, B. (2017).
\newblock Evaluating and improving fault localization.
\newblock In {\em 2017 IEEE/ACM 39th International Conference on Software Engineering (ICSE)\/}, pages 609--620.

\bibitem[Rafi {\em et~al.}(2024)Rafi, Pacheco, Chen, Yang, and Chen]{sbest}
Rafi, M.~N., Pacheco, L. B.~S., Chen, A.~R., Yang, J., and Chen, T.-H.~P. (2024).
\newblock Leveraging stack traces for spectrum-based fault localization in the absence of failing tests.
\newblock \url{https://zenodo.org/records/11062413}.
\newblock Zenodo.

\bibitem[Rafi {\em et~al.}(2025)Rafi, Chen, Chen, and Wang]{chen2023back}
Rafi, M.~N., Chen, A.~R., Chen, T.-H.~P., and Wang, S. (2025).
\newblock Revisiting defects4j for fault localization in diverse development scenarios.
\newblock In {\em 2025 IEEE/ACM 22nd International Conference on Mining Software Repositories (MSR)\/}, pages 63--75.

\bibitem[Saha {\em et~al.}(2013)Saha, Lease, Khurshid, and Perry]{saha2013improving}
Saha, R.~K., Lease, M., Khurshid, S., and Perry, D.~E. (2013).
\newblock Improving bug localization using structured information retrieval.
\newblock In {\em 2013 28th IEEE/ACM International Conference on Automated Software Engineering (ASE)\/}, pages 345--355. IEEE.

\bibitem[Schroter {\em et~al.}(2010)Schroter, Schröter, Bettenburg, and Premraj]{schroter2010stack}
Schroter, A., Schröter, A., Bettenburg, N., and Premraj, R. (2010).
\newblock Do stack traces help developers fix bugs?
\newblock In {\em 2010 7th IEEE Working Conference on Mining Software Repositories (MSR 2010)\/}, pages 118--121.

\bibitem[Shah {\em et~al.}(2008)Shah, G{\"o}rg, and Harrold]{shah2008developers}
Shah, H., G{\"o}rg, C., and Harrold, M.~J. (2008).
\newblock Why do developers neglect exception handling?
\newblock In {\em Proceedings of the 4th international workshop on Exception handling\/}, pages 62--68.

\bibitem[Silva {\em et~al.}(2021)Silva {\em et~al.}]{silva-etal-2021-flacoco}
Silva, A. {\em et~al.} (2021).
\newblock Flacoco: Fault localization for java based on industry-grade coverage.
\newblock {\em arXiv preprint arXiv:2111.12513\/}.

\bibitem[Sohn and Yoo(2017)Sohn and Yoo]{sohn2017fluccs}
Sohn, J. and Yoo, S. (2017).
\newblock Fluccs: Using code and change metrics to improve fault localization.
\newblock In {\em Proceedings of the 26th ACM SIGSOFT International Symposium on Software Testing and Analysis\/}, pages 273--283.

\bibitem[Srivastava(2008)Srivastava]{srivastava2008test}
Srivastava, P.~R. (2008).
\newblock Test case prioritization.
\newblock {\em Journal of Theoretical \& Applied Information Technology\/}, {\bf 4}(3).

\bibitem[Vancsics {\em et~al.}(2021)Vancsics, Horvath, Szatmari, and Beszedes]{callfreq}
Vancsics, B., Horvath, F., Szatmari, A., and Beszedes, A. (2021).
\newblock Call frequency-based fault localization.
\newblock In {\em 2021 IEEE International Conference on Software Analysis, Evolution and Reengineering (SANER)\/}, pages 365--376.

\bibitem[Wang {\em et~al.}(2022)Wang, Li, Liu, Chen, Paul, Cai, and Fan]{9759520}
Wang, H., Li, Z., Liu, Y., Chen, X., Paul, D., Cai, Y., and Fan, L. (2022).
\newblock Can higher-order mutants improve the performance of mutation-based fault localization?
\newblock {\em IEEE Transactions on Reliability\/}, {\bf 71}(2), 1157--1173.

\bibitem[Wang {\em et~al.}(2020)Wang, Yao, Tong, Huo, Li, Xu, and Lu]{wang2020enhancing}
Wang, Y., Yao, Y., Tong, H., Huo, X., Li, M., Xu, F., and Lu, J. (2020).
\newblock Enhancing supervised bug localization with metadata and stack-trace.
\newblock {\em Knowledge and Information Systems\/}, {\bf 62}, 2461--2484.

\bibitem[Wen {\em et~al.}(2021)Wen, Chen, Tian, Wu, Hao, Han, and Cheung]{8873606}
Wen, M., Chen, J., Tian, Y., Wu, R., Hao, D., Han, S., and Cheung, S.-C. (2021).
\newblock Historical spectrum based fault localization.
\newblock {\em IEEE Transactions on Software Engineering\/}, {\bf 47}(11), 2348--2368.

\bibitem[Wong {\em et~al.}(2014)Wong, Xiong, Zhang, Hao, Zhang, and Mei]{wong2014boosting}
Wong, C.-P., Xiong, Y., Zhang, H., Hao, D., Zhang, L., and Mei, H. (2014).
\newblock Boosting bug-report-oriented fault localization with segmentation and stack-trace analysis.
\newblock In {\em 2014 IEEE International Conference on Software Maintenance and Evolution\/}, pages 181--190.

\bibitem[Wong {\em et~al.}(2013)Wong, Debroy, Gao, and Li]{wong2013dstar}
Wong, W.~E., Debroy, V., Gao, R., and Li, Y. (2013).
\newblock The dstar method for effective software fault localization.
\newblock {\em IEEE Transactions on Reliability\/}, {\bf 63}(1), 290--308.

\bibitem[Wu {\em et~al.}(2014)Wu, Zhang, Cheung, and Kim]{wu2014crashlocator}
Wu, R., Zhang, H., Cheung, S.-C., and Kim, S. (2014).
\newblock Crashlocator: Locating crashing faults based on crash stacks.
\newblock In {\em Proceedings of the 2014 International Symposium on Software Testing and Analysis\/}, pages 204--214.

\bibitem[Xia {\em et~al.}(2016)Xia, Bao, Lo, and Li]{xia2016automated}
Xia, X., Bao, L., Lo, D., and Li, S. (2016).
\newblock “automated debugging considered harmful” considered harmful: A user study revisiting the usefulness of spectra-based fault localization techniques with professionals using real bugs from large systems.
\newblock In {\em 2016 IEEE International Conference on Software Maintenance and Evolution (ICSME)\/}, pages 267--278. IEEE.

\bibitem[Yang {\em et~al.}(2024)Yang, Le~Goues, Martins, and Hellendoorn]{yang2024large}
Yang, A.~Z., Le~Goues, C., Martins, R., and Hellendoorn, V. (2024).
\newblock Large language models for test-free fault localization.
\newblock In {\em Proceedings of the 46th IEEE/ACM International Conference on Software Engineering\/}, pages 1--12.

\bibitem[Ye {\em et~al.}(2022)Ye, Martinez, and Monperrus]{ye2022neural}
Ye, H., Martinez, M., and Monperrus, M. (2022).
\newblock Neural program repair with execution-based backpropagation.
\newblock In {\em Proceedings of the 44th International Conference on Software Engineering\/}, pages 1506--1518.

\bibitem[Zhang {\em et~al.}(2021)Zhang, Liu, Kim, Li, Liu, Klein, and Bissyandé]{9609138}
Zhang, J., Liu, K., Kim, D., Li, L., Liu, Z., Klein, J., and Bissyandé, T.~F. (2021).
\newblock Revisiting test cases to boost generate-and-validate program repair.
\newblock In {\em 2021 IEEE International Conference on Software Maintenance and Evolution (ICSME)\/}, pages 35--46.

\bibitem[Zhang {\em et~al.}(2024)Zhang, Ruan, Fan, and Roychoudhury]{zhang2024autocoderover}
Zhang, Y., Ruan, H., Fan, Z., and Roychoudhury, A. (2024).
\newblock Autocoderover: Autonomous program improvement.
\newblock In {\em Proceedings of the 33rd ACM SIGSOFT International Symposium on Software Testing and Analysis\/}, pages 1592--1604.

\bibitem[Zhou {\em et~al.}(2012)Zhou, Zhang, and Lo]{zhou2012should}
Zhou, J., Zhang, H., and Lo, D. (2012).
\newblock Where should the bugs be fixed? more accurate information retrieval-based bug localization based on bug reports.
\newblock In {\em 2012 34th International conference on software engineering (ICSE)\/}, pages 14--24. IEEE.

\bibitem[Zou {\em et~al.}(2021)Zou, Liang, Xiong, Ernst, and Zhang]{zou2021flfamily}
Zou, D., Liang, J., Xiong, Y., Ernst, M.~D., and Zhang, L. (2021).
\newblock An empirical study of fault localization families and their combinations.
\newblock {\em IEEE Transactions on Software Engineering\/}, {\bf 47}(2), 332--347.

\end{thebibliography}

\end{document}